# RHEOLOGY OF WORMLIKE MICELLES : EQUILIBRIUM PROPERTIES AND SHEAR BANDING TRANSITION

Jean-François Berret
*Complex Fluids Laboratory, Unité Mixte de Recherche Rhodia-CNRS n° 166,*
*259 Prospect Plains Road CN 7500, Cranbury NJ 08512 USA*

**Abstract**
We review the experimental and theoretical results obtained during the past decade on the structure and rheology of wormlike micellar solutions. We focus on the linear and nonlinear viscoelasticity and emphasize the analogies with polymers. Based on a comprehensive survey of surfactant systems, the present study shows the existence of standard rheological behaviors for semidilute and concentrated solutions. One feature of this behavior is a shear banding transition associated with a stress plateau in the nonlinear mechanical response. For concentrated solutions, we show that in the plateau region the shear bands are isotropic and nematic.

## INTRODUCTION

Wormlike micelles are elongated and semiflexible aggregates resulting from the self-assembly of surfactant molecules in aqueous solutions. In the general context of complex fluids, wormlike micelles have received considerable attention from theoreticians and experimentalists during the past decade [1-8]. One reason for this interest is due to their remarkable rheological properties. When micelles grow and become wormlike, the aggregates are much like polymers, and as polymers they entangle above a critical concentration. The aqueous solutions then become viscoelastic. Quantitative rheological measurements show that this viscoelasticity is characterized by a single relaxation time, a property which is rather unusual for fluids with complex microstructures [9]. This rule is indeed so general that it is now admitted that a single relaxation time in the linear mechanical response is a strong indication of the wormlike character of self-assembled structures. This property has incited several groups to use wormlike micelles as reference systems for the testing of new experimental techniques [10-13]. Wormlike micelles are also considered as models for polymers because of their nonlinear rheological properties. When submitted to steady shear, these viscoelastic fluids undergo a shear banding transition, which is associated with a plateau in the stress *versus* shear rate curve. The shear banding transition is a transition between an homogeneous and a non homogeneous state of flow, the latter being characterized by a "separation" of the fluid into macroscopic regions (bands) of different shear rates. This type of transitions is thought to be analogous in nature to the instability found in extrusion of polymer melts at high temperature. Since the first reports on shear banding in micellar solutions [14-16], non-homogeneous flows were observed in many other complex systems. A final but also an important reason to study wormlike micelles is the widespread range of their applications in today's life. Viscoelastic surfactant phases are already used in oil fields as fracturing fluids, in hydrodynamic engineering as drag reducing agents and in many home and personal care products [8,17].

This review is organized as follows. In the first part, we investigate the equilibrium properties and emphasize three properties that are crucial for the rheology : the dynamics of growth of the aggregates, the role of the flexibility on the phase behavior and the relaxation dynamic in the viscoelastic regime. Each aspect is illustrated by data and references taken from the recent literature. We also provide several results that have not yet been published, as for instance on the determination of the persistence length. Scaling laws express the specific dependences of some physical quantities, such as the viscosity or the scattering with the concentration. A critical analysis of the scaling properties found in wormlike micelles is proposed. The second part of the review deals with nonlinear rheological properties of semidilute and concentrated phases of wormlike micelles. Based on a broad bibliographic survey, our approach tends to demonstrate the existence of a "standard" behavior for all these systems. The characteristics of the "standard" behavior are a Maxwellian behavior of the linear viscoelasticity and a shear banding transition in the nonlinear response. We explore the diversity of the features of shear banding in micelles and show that for concentrated systems the instability is associated with a thermodynamic transition between an isotropic and a nematic state.

## 1.    EQUILIBRIUM PROPERTIES

### 1.1 – Theoretical Background

Wormlike micelles are elongated and semiflexible aggregates resulting from the self-assembly of surfactant molecules in aqueous solutions. The growth and stability of micellar aggregates are in general described in terms of packing of the surfactant molecules, or equivalently in terms of curvature of the water/hydrocarbon interface. Above the critical micellar concentration (c.m.c.), spherical micelles form spontaneously and their size is related to that of the amphiphilic molecules. The kinetics of the micelle formation and breakdown above the c.m.c. is based multiple equilibria, in which the micelles grow or shrink by stepwise incorporations or dissociations of monomers. This equilibrium is known already for some time and its description can be found in reviews [18-20] and textbooks [4,21,22]. Several parameters can be adjusted in order to modify the curvature of the water/hydrocarbon interface and to favor a change of morphology. These parameters are the surfactant concentration, the ionic strength, the temperature etc. For surfactant showing preferentially the cylindrical aggregation, the end-cap energy E denotes the excess in packing energy (between a spherical and a cylindrical environment) for the molecules located in the two hemispherical end-caps. In the following we derive the growth laws for neutral and polyelectrolyte micelles.

*Neutral Micelles* : The end-cap energy E is here equivalent to the scission energy necessary to create two new chain ends. For a dispersion of micelles of length L and molecular weight distribution c(L), the minimization of a free energy that takes into account the end-cap energy and the translational entropy yields for the average micellar length [4,22,23] :

$$\overline{L} = \frac{2}{n_0} c^{1/2} \exp\left(\frac{E}{2k_B T}\right) \qquad (1)$$

where $n_0$ is the number of surfactant per unit length of the linear aggregate. $n_0$ (in $Å^{-1}$) is of the order of unity. End-cap energy have been determined from temperature jump measurements and they are in the range $20 - 30\ k_B T$ [1,24]. Note that the same type of reasoning for two-dimensional aggregation leads to the conclusion that disc-like aggregates can only exist as infinite bilayers. The distribution in length c(L) is broad and given by :

$$c(L) = \frac{c}{\overline{L}^2} \exp\left(-\frac{L}{\overline{L}}\right) \qquad (2)$$

where c(L)dL denotes the number density of chains of length comprised between L and L + dL. The exponential distribution in Eq. 2 corresponds to an index of polydispersity of 2.

*Polyelectrolyte Micelles* : Electrostatically charged micelles are made from ionic surfactants and the solutions are prepared with no added salt. For these aggregates, point charges located at the hydrocarbon/water interface modify the micellar growth and length distribution. This effective charge of the aggregate arises actually from the incomplete compensation of the surfactant charges by the counterions. MacKintosh and coworkers have proposed a model to demonstrate that the electrostatic interactions reduce the scission energy and favor the breaking of micelles [25,26]. For polyelectrolyte micelles, the end-cap energy is not equivalent to the scission energy. The electrostatic contribution to the free energy results in a broad dilute regime. There, the micelles are rather monodisperse and their length is only very slowly increasing with concentration. The overlap concentration c* between the dilute and semidilute regimes depends on the effective linear charge density noted $\nu$, through the relationship :

$$c^* = \left( \frac{k_B T \ell_B R_C \nu^2}{E} \right)^2 \qquad (3)$$

where $\ell_B$ is the Bjerrum length ($\ell_B = 7.15$ Å in water) and $R_C$ the radius of the cylinder. Above the overlap threshold, which also corresponds to the screening of the electrostatic interactions by the counterions, the micelles grow more rapidly according to :

$$\bar{L} = \frac{2}{n_0} c^{1/2} \exp\left( \frac{E}{2 k_B T} \left[ 1 - \left( c^* / c \right)^{1/2} \right] \right) \qquad (4)$$

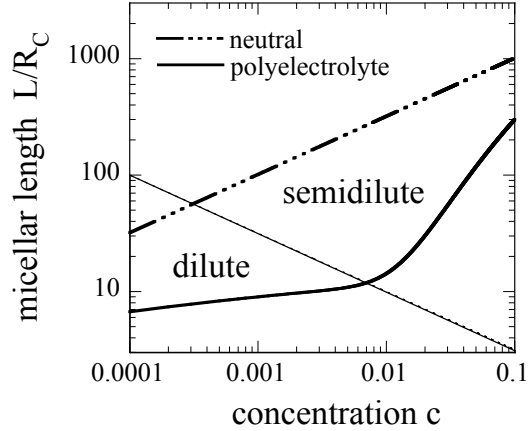

**Figure 1 :** Predictions for the average micellar length of neutral and polyelectrolyte wormlike micelles, as function of the surfactant concentration. The dash-dotted and continuous lines are calculated according to Eqs. 1 and 4 respectively, with parameters taken from [25,26] : $E = 20$ $k_B T$, $\ell_B R_C \nu^2 = 1$ and $n_0 = 1/2$.

Fig. 1 displays the growth laws for neutral and charged micelles, according to Eq. 1 and Eq. 4 respectively. In Fig. 1, the ratio $\bar{L}/R_C$ is plotted as function of c, the surfactant concentration and using parameters indicated in the caption [25,26]. The predictions for the dynamics of growth of neutral and polyelectrolyte micelles agree qualitatively with experiments. For instance, it is known that the addition of salt to a solution of polyelectrolyte micelles can result in a strong increase of viscosity, the solution passing from a viscous fluid to a viscoelastic gel. In Fig. 1, this change corresponds to a transition at a fixed concentration from the bottom curve (polyelectrolyte, dilute regime) to the top curve (neutral, semidilute regime). Quantitatively however and except for few systems [27], the analytic forms of the growth laws for neutral and polyelectrolyte micelles have not been found systematically. This issue will be discussed at the end of the first part.

## 1.2 – Physical Chemistry of Wormlike Micelles and Related Systems

Wormlike micelles can form spontaneously at ambient temperature using cationic surfactants with e.g. 16 carbon atoms in the aliphatic chain. This is the case for cetyltrimethylammonium bromide (CTAB) [28-30] and cetylpyridinium bromide (CPBr) [31]. As we have seen previously, because of electrostatics the transition between spherical to cylindrical aggregates occurs at relatively high surfactant concentrations (Fig. 1). The growth of the aggregates can be promoted however if cosurfactants or other low molecular weight additives are incorporated to the solutions. These additives are short alcohol chains, strongly binding counterions, oppositely charged surfactants etc. We review below the different classes of surfactants and cosurfactants/additives which form such structures. A list of the most common surfactants and counterions known to form wormlike micelles is given in Table I. Table I also provides the chemical formula of these molecules and their abbreviations. During the last decade, a number of other low molecular weight molecules were found to associate in solutions into elongated structures. Because in many cases their rheology is close to that of the surfactant micelles, we have included them to the listing below.

**A – Surfactant and simple salt.** The addition of simple salts such as sodium chloride (NaCl) or potassium bromide (KBr) to ionic surfactant solutions results in the screening of the electrostatic interactions between the charges, and thus in the growth of the aggregates. The archetype system of class **A** is CTAB with KBr [5,20,24,32-35]. Other well-known examples are sodium dodecyl sulfate (SDS) with monovalent [27,36-41] or multivalent counterions [42,43].

**B – Surfactant and cosurfactant**, where the cosurfactant is a short alcohol chain. Classical examples are the ternary systems sodium alkylsulfate-decanol-water (SdS-Dec [44-47] and SDS-Dec [48-50], see table I) and cetylpyridinium chloride-hexanol-brine (CPCl-Hex) [51-55]. In these systems, the ratio between the alcohol and surfactant concentrations controls the polymorphism of the self-assembly. The theoretical arguments developed in Section 1.1 for neutral chains should apply to this class, namely those for which the cylindrical aggregates are intermediate structures between spheres and bilayers.

**C – Surfactant and strongly binding counterion.** Strongly binding counterions are small molecules of opposite charge with respect to that of the surfactant. They are sometimes called hydrotopes. Well-known examples of hydrotopes are salicylate, tosilate and chlorobenzoate counterions, which all contain an aromatic phenyl group (see Table I). CTAB and CPCl with sodium salicylate (NaSal) have been probably the most studied micellar systems during the last two decades [3,56-70]. Contrary to simple salts (class **A**), a large proportion of these counterions (~ 80 %) is assumed to be incorporated into the micelles. It was found that in CPCl-NaSal, long wormlike micelles are immediately formed at the c.m.c. (0.04 wt. %), without passing through an intermediate spherical morphology [68,69,71].

| name | developed formula | counterions | Abbr. |
|---|---|---|---|
| decyl sulfate | $(C_{10}H_{21})$-$SO_3^-$ | $Na^+$ | SdS |
| dodecyl-trimethylammonium | $(C_{12}H_{25})$-$N^+$-$(CH_3)_3$ | $Br^-$ | DTAB |
| dodecyl sulfate | $(C_{12}H_{25})$-$SO_3^-$ | $Na^+$ | SDS |
| tetradecyl-trimethylammonium | $(C_{14}H_{29})$-$N^+$-$(CH_3)_3$ | $Br^-$ | TTAB |
| cetyl-trimethylammonium | $(C_{16}H_{33})$-$N^+$-$(CH_3)_3$ | $Br^-$ $Cl^-$ $CH_3$-$(C_6H_4)$-$SO_3^-$ $Cl$-$(C_6H_4)$-$COO^-$ | CTAB CTAC CTAT CTAClBz |
| cetylpyridinium | $(C_{16}H_{33})$-$(C_5H_5)$-$N^+$ | $Br^-$ $Cl^-$ $ClO_3^-$ $OH$-$(C_6H_4)$-$COO^-$ | CPBr CPCl CPClO_3 CPSal |
| dodecyl-benzenesulfonate | $(C_{12}H_{25})$-$(C_4H_6)$-$SO_3^-$ | $Na^+$ | SDBS |
| tetradecyl-dimethylamine oxide | $(C_{14}H_{29})$-$N^+$-$(CH_3)_2$-$OH$ | $Cl^-$ | C14DMAO |
| hexadecyloctyl-dimethylammonium | $(C_{16}H_{33})$-$(C_8H_{17})$-$N^+$-$(CH_3)_2$ | $Br^-$ | C18-C8DAB |

**Table I** : Most common surfactants and counterions known to form wormlike micelles in water. The counterions in the third column with developed formula $OH$-$(C_6H_4)$-$COO^-$, $CH_3$-$(C_6H_4)$-$SO_3^-$ and $Cl$-$(C_6H_4)$-$COO^-$ are termed salicylate (Sal), toluenesulfonate (or tosilate, abbreviated as T or Tos) and chlorobenzoate (ClBz), respectively.

**D – Amphoteric surfactant**. Amphoteric surfactants are surface active molecules that contain positive and negative charges in the head group. Betaine-type molecules with quaternary ammonium and carboxylate groups are the representatives of this class. They associate at low concentrations and aqueous solutions exhibit strong gel-like properties. These properties are attributed to the existence of an entangled network of micelles [72,73].

**E – Gemini surfactants and surfactant oligomers.** The covalent linking of amphiphilic moieties at the level of the head group yields to Gemini surfactants and surfactant oligomers (for a review, see [74]). In aqueous solutions, these molecules present a broad polymorphism of aggregation [75-79]. Gemini surfactants are one of the rare examples for which cylindrical micelles close on themselves spontaneously, forming loops or rings. This properties has been attributed to large end-cap energies [80].

**F − Cationic and anionic mixtures.** Oppositely charged surfactants have shown synergistic enhancements of rheological properties, and notably through the formation of mixed wormlike micelles. The growth of the micelles is assumed to arise from the charge neutralization of the surface potential (as in **C**) and from the related increase of the ionic strength (as in **A**). Recent examples studied are the mixtures of sodium dodecylsulfate (SDS) and dodecyltrimethylammonium bromide (DTAB) [81,82], or the mixtures made from cetyltrimethylammonium tosilate and sodium dodecyl benzenesulfonate [83,84].

**G − Reverse micelles in organic solvent**. Lecithin is a phospholipid that is a major component in the lipid matrix of biological membranes. When dissolved in a nonpolar solvent (an alkane), reverse spherical micelles are formed. The addition of small amount of water triggers the growth of aggregates. The rheology of these solutions was shown to be similar to that of direct wormlike micellar solutions [85-88].

**H − Metallic salts in organic solvent.** In apolar solvent, some metallic salts such as the copper tetracarboxilate associate into elongated wires. The wires are built with only one molecule in the cross section. As in surfactant micelles, the aggregation number can be very large and the wires are described as long and semiflexible colloids. With increasing concentrations, solutions become strong viscoelastic jellies, with the same signatures as those of surfactant systems [89-91].

**I − Block copolymers.** Cylindrical self-assembly have also been reported in aqueous solutions of low molecular weight block copolymers [92-94]. The system investigated in [92,93] is poly(ethylene oxide)-poly(butadiene) with a weight fraction around 50 % for the first block. "Giant" micelles have been found by cryo-transmission electron microscopy and the cylindrical morphology of the poly(butadiene) core was confirmed by neutron scattering.

## 1.3    Flexibility of Wormlike Micelles

### 1.3.1    Persistence Length

The persistent character of the micellar chains is a fundamental feature of wormlike micelles. The flexibility of the chains determines the equilibrium conformations in good solvent, as well as the phase behavior of the solutions at high concentrations. Fig. 2 shows the segment of a cylindrical micelle that is uniformly bent. The vector $\mathbf{u}(s)$ is tangent to the micellar axis and s denotes the curvilinear length. The bending free energy per unit of s is proportional to the product $\kappa C^2$, where $\kappa$ is the bending modulus of the chain and C its curvature. For surfactants, Ben-Shaul and coworkers have shown that this free energy has a molecular origin [95]. Contributions arising from the repulsions between head groups, from the hydrocarbon-water interfacial energy and from the conformations of the aliphatic chains in the core should be taken into account in the calculation of κ.

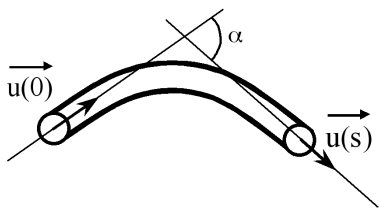

**Figure 2** : Representation of a segment of a wormlike micelle. The angular correlation function of the vectors **u**(s) tangent to the micellar axis is used to define the persistence length b of the polymer-like aggregate (Eq. 5). For wormlike micelles, b is of the order of 400 Å.

From a geometrical point of view, the persistence length, noted b measures the decay length of the angular correlation along the micellar axis and can be defined through the equality :

$$\langle \mathbf{u}(0).\mathbf{u}(s) \rangle = \exp(-s/b) \tag{5}$$

where the brackets denote the average over the orientational distribution of micellar segments. From a thermodynamical point of view, b is associated to the bending modulus through the relationship $b = \kappa/k_B T$. b is the length over which the thermal fluctuations are able to cancel the orientational coherence of the chain. The first experimental determination of the persistence length in wormlike micellar solutions was carried out in 1980 by Porte and coworkers, combining dynamic light scattering and magnetic birefringence experiments. The system placed under scrutiny was cetylpyridinium bromide (CPBr) with sodium bromide (class **A**) [96,97]. For this system, the saturation of the magnetic field-induced birefringence observed with increasing surfactant concentration (and in a range where the hydrodynamic radius was still growing) was interpreted as an evidence of the independent alignment of segments of length b. b was estimated of the order of 200 Å [53,98,99]. It has been recognized since then that elongated cylindrical micelles of surfactants must be described as semiflexible chains. In this section, we show how scattering techniques and in particular small-angle neutron scattering can be used to determine the persistence length of wormlike micelles.

### 1.3.2    Scattering Function of a Persistent Micellar Chain

Small-Angle Neutron Scattering (SANS) is an ideal tool for studying surfactant polymorphism because of the large scattering contrast resulting from the use of hydrogenated surfactants and deuterated water as a solvent.

For wormlike micelles, SANS presents a second advantage. The radius $R_C$ and the persistence length b fall in the wave-vector range covered by this technique (q = $10^{-3}$ Å$^{-1}$ – 0.4 Å$^{-1}$) and their respective signatures are well separated. The average length of micelles can also be obtained, but the wave-vector limitations restrict this determination to values below ~ 1000 Å [27,41,81]. Light scattering [100] and ultra small-angle neutron scattering are then useful complementary techniques. Our goal here is to derive the scattering function of a persistent chain as function of the wave-vector q and to show how it compares to experimental data on one example. The scattering cross section of semiflexible chains is usually derived from asymptotic behaviors [101]. The two wave-vectors ranges relevant for SANS are qb < 1 and qb > 1. In the following, we referred to them as low and high wave-vectors ranges, respectively. We assume furthermore the inequality $R_C$ < b < $\overline{L}$.

**qb > 1** (high q-range) : In this range, micelles appear to neutrons as dispersed and disordered rods of lengths shorter than b, and the scattering cross section dσ(q)/dΩ reads :

$$\frac{d\sigma}{d\Omega}\left(q, qb > 1\right) = 4\pi^3 \Delta\rho^2 \mathcal{L} R_C^4 \frac{1}{q} \left(\frac{J_1(qR_C)}{qR_C}\right)^2 \qquad (6)$$

where $J_1$ denotes the first-order Bessel function, $\Delta\rho$ the difference in scattering length densities between the aggregates and the solvent. For surfactant micelles with core radius $R_C$ of 20 Å, the range of validity of Eq. 6 goes from 0.01 to 0.4 Å$^{-1}$. Note that the average micellar length $\overline{L}$ does not enter Eq. 6 explicitly. Instead, there is the total micellar length per unit volume $\mathcal{L}$ in prefactor. For a solution prepared at concentration **c** with a surfactant of molecular weight $M_W$, the ratio $c\mathcal{N}_A/M_W\mathcal{L}$ is the number of surfactants per unit length of micelles, noted $n_0$ in Eqs. 1 and 4. This ratio can thus be determined accurately by SANS and it is of the order of $n_0$ = 2 molecules/Å [41,102,103]. For $qR_C$ < 1, the expression in parenthesis in Eq. 6 tends to 1/2 and the scattering function varies as q$^{-1}$.

**qb < 1** (low q-range) : At low wave-vectors, the scattering probes the self-avoiding random walk configuration of the chains. The intensity decreases with a power law with exponent which is –2 for Gaussian chains and –5/3 for chains in good solvent. The flexibility of micellar aggregates manifests itself as a cross-over between a q$^{-1}$- and a q$^{-2}$ (or q$^{-5/3}$)-behaviors, this cross-over occurring in the range q ~ 1/b. This reasoning is also valid for the semidilute regime, if the curvilinear length $L_e$ between entanglements is much larger than the persistence length b. $L_e$ is related to the mesh size ξ of the network through the equation $L_e \sim \xi^{5/3} b^{-2/3}$ [104].

The previous ideas are illustrated in Fig. 3. Here, the form factor of a single chain of length L = 4 µm, of persistence length b = 400 Å and of radius $R_C$ = 20 Å is shown as function of the wave-vector. The calculation is made using the analytical expressions provided by Brûlet *et al.* and by Pedersen and Schurtenberger without excluded volume interactions [105,106]. In the inset of Fig. 3, the transition between the $q^{-1}$ and a $q^{-2}$ regimes is emphasized. This transition is smooth and continuous.

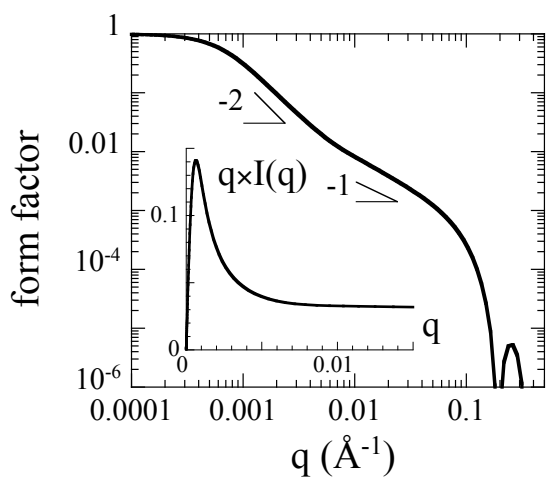

**Figure 3** : Scattering form factor of a wormlike micellar aggregate of total curvilinear length L = 4 µm, of persistence length b = 400 Å and of radius $R_C$ = 20 Å. Calculations were made using the Pedersen and Schurtenberger model [106]. The cross-over between the $q^{-1}$ and $q^{-2}$ regime is emphasized in the inset (Holtzer representation, q×I(q) *versus* q).

### 1.3.3    Comparison with Small-Angle Neutron Scattering

In Fig. 4 are displayed data obtained on several CTAB solutions prepared at c = 0.2 wt. % and using increasing amounts of sodium salicylate (NaSal). In this figure and in the sequel of the review, the parameter R denotes the molar ratio between the two components, R = [Sal-]/[CTA+]. At R = 0, *i.e.* for pure CTAB solution, the scattering cross section dσ(q)/dΩ is low and indicative of interacting spherical micelles. At R = 0.36, the neutron spectrum is modified and exhibits an increase in scattering at low q. At R = 1.6 and above, all sets of data display the $q^{-1}$-behavior characteristic of elongated aggregates. Neutron scattering thus provides a clear illustration of the micellar growth by addition of strongly binding counterions. At this concentration and values of R, the CTAB-NaSal solutions are in the slightly entangled regime and the curvilinear length between entanglements is much larger than the persistence length. We have compared these data to the predicted cross section for semiflexible aggregates [105,106]. In Fig. 5, the product q×dσ(q)/dΩ is shown in double logarithmic scales for the system CTAB-NaSal at molar ratios 7.3 and 36.4. This representation, also called Holtzer representation emphasizes the transition between the $q^{-1}$ and $q^{-2}$-asymptotic regimes. The data are those of Fig. 4. The continuous lines through the data are calculated using b = 380 Å and b = 360 Å, respectively. Persistence lengths of the order of 400 Å are typical for neutral wormlike micelles, and have been found repeatedly in the recent literature [27,53,98,100,102,107]. For polyelectrolyte micelles, the total persistence length can be larger than the above characteristic values [108,109]. In conclusion, the persistence length of micellar aggregates is of the order of 400 Å, a value that is comprised between that of synthetic polymers [104] and that of biological molecules, such as DNA, actin and tubuline [110,111].

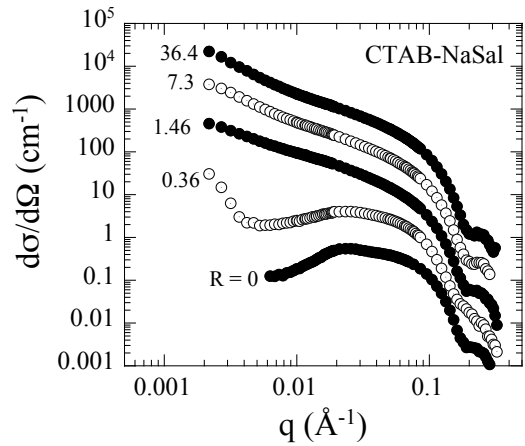

**Figure 4** : Neutron scattering cross section $d\sigma(q)/d\Omega$ *versus* q for CTAB-Sal micelles for molar ratios R = [Sal⁻]/[CTA⁺] between R = 0 and R = 36.4. The CTAB concentration is c = 0.2 wt. % (5.5 mmol/l) for all solutions. The scattering curves have been shifted by a factor 5 with respect to each other. The increase of the scattering at low q with increasing R is interpreted as the evidence of the micellar growth.

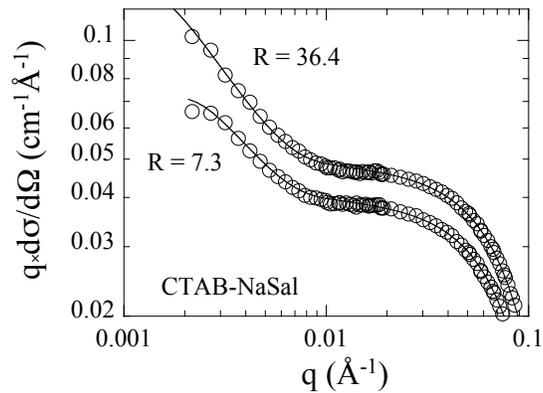

**Figure 5** : $q \times d\sigma(q)/d\Omega$ *versus* q for CTAB-NaSal wormlike micelles at surfactant concentration c = 0.2 wt. % and molar ratios R = 7.3 and R = 36.4. The experimental data are those of Fig. 4. The continuous lines are calculated according to the model described by Pedersen and Schutenberger [106]. The adjustable parameters are the micellar radius $R_C$, the persistence length b and the average micellar length between entanglements. For R = 7.3, b = 380 Å and L/b = 15 and for R = 36.4, b = 360 Å and L/b = 40 (with $R_C$ = 20 Å for both). The intensity for R = 36.4 has been shifted by a factor 1.3.

## 1.4    Phase Behavior

At high surfactant concentration, the average contour length of wormlike micelles can be of the order of microns *i.e.* much larger than the persistence length b, *i.e.* L >> b >> $R_C$. Semenov and Kokhlov have shown than for semiflexible chains obeying the last inequalities long-range orientational order appears with increasing volume fraction as a result of a first-order phase transition [112-114]. The transition is between an isotropic disordered phase and an orientationally ordered nematic phase. Based on the approach developed for stiff chains [115], and assuming only steric interactions between chains, the values of the phase boundaries between the ordered and disordered states, as well as the order parameter of the nematic phase were predicted. These boundaries are denoted $c_{I\text{-}N}$ and $c_N$ and are defined as follows. For $c < c_{I\text{-}N}$, the solutions are isotropic, at $c > c_N$ they are nematic, and for $c_{I\text{-}N} < c < c_N$, they separate into an isotropic and a nematic phase. For semiflexible chains, Semenov and Khokhlov have found :

$$c_{I\text{-}N} = 10.48 \, \frac{R_C}{b}, \quad c_N = 11.39 \, \frac{R_C}{b} \quad \text{and} \quad w = \frac{c_N}{c_{I\text{-}N}} - 1 = 0.09 \quad (7)$$

The remarkable result in Eqs. 7 is that the phase boundaries depend only on the ratio between the radius and the persistence length. For rigid rods, the same type of equations was obtained, the persistence length being replaced by the average length of the rods. For rods, the prefactors are also different from those in Eqs. 7 [112,114,115].

On the experimental side, nematic phases of wormlike micelles have been found systematically at high surfactant concentrations, between 20 and 50 wt. % [44,46-48,50,52,71,116-120]. Examples of phase diagrams reported recently in the literature are shown in Figs. 6. Three ternary systems are presented, which are cetylpyridinium chloride–hexanol–water (at 0.2 M NaCl) [52,119,121], cetylpyridinium chloride–sodium salicylate–water (at 0.5 M NaCl) [122] and sodium dodecyl sulfate–decanol–water [49,50,120]. In Fig. 6d, the binary system cetyltrimethylammonium bromide–water is shown and the phases are displayed in a temperature-concentration diagram [30,117,123]. Note for CPCl-Sal the presence of two nematic "islands", one at salicylate concentration around 5 wt. % and one around 20 wt. %. All four systems show that at constant ratio with respect to the alcohol or hydrotope content, there is the same sequence of thermodynamically stable phases : isotropic, nematic and hexagonal. The packing of semiflexible cylinders thus leads to mesophases with orientational and translational orders. For the four systems, the limits of phases $c_{I-N}$ and $c_N$ (varying between 20 and 45 wt. %) compare well with the theoretical values predicted for semiflexible chains with radius 20 Å and persistence length 400 Å (as previously determined). Eq. 7 gives values for $c_{I-N}$ and $c_N$ around 50 wt. %. The deviations observed for SDS-Dec and CTAB are the largest and they could be due to additional electrostatic interactions.

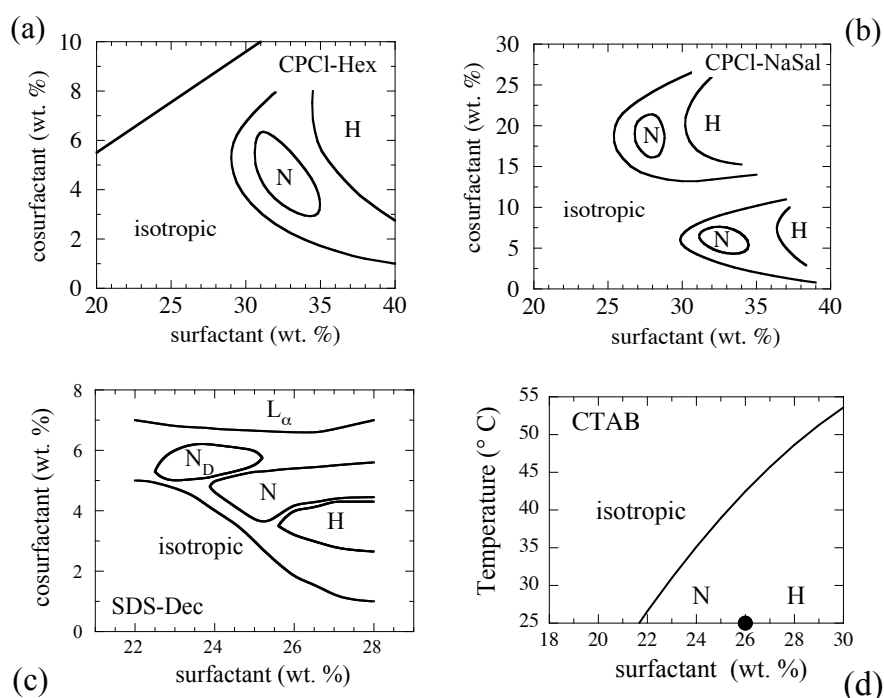

**Figure 6** : Phase diagrams of four wormlike micellar systems : a) cetylpyridinium chloride, hexanol and water (0.2 M NaCl) [121]; b) Cetylpyridinium chloride, sodium salicylate and water (0.5 M NaCl) [122]; c) sodium dodecylsulfate, decanol and water [49,50,120]; d) cetyltrimethylammonium bromide and water [30,124]. For the diagrams (a), (b) and (c), the cosurfactant concentration is in ordinate. For diagram (d), it is the temperature. The phase boundary between the nematic and hexagonal phases in CTAB-H$_2$O is not known. According to Fontell *et al.*, the transition occurs at c = 26 wt. % at room temperature (T = 25° C) [30]. At fixed cosurfactant/surfactant ratios or fixed temperature, similar sequences of phases are observed with increasing concentrations: isotropic (entangled), nematic (N) and hexagonal (H). N$_D$ denotes nematic discotic and L$_\alpha$ lamellar.

## 1.5    Linear Rheology and Scaling

That surfactant solutions can be strongly viscoelastic was noticed by several authors as early as 1950's. Nash for instance identified the role of additives such as naphtalene derivatives in the onset of viscoelasticity in CTAB solutions [56]. One intriguing result was that the viscoelasticity of the solution was showing up well below the c.m.c. of the surfactant. Some years later, Gravsholt and coworkers recognized that other types of additives, such as salicylate or chlorobenzoate counterions are actually solubilized by the micelles, lowering then the c.m.c. of the surfactant [59,125]. It was suggested by the same authors that the viscoelasticity had the same physical origin than that of polymer solutions and melts, namely entanglements and reptation [104]. The picture proposed in late 1970's was that of a network of entangled rod-like micelles.

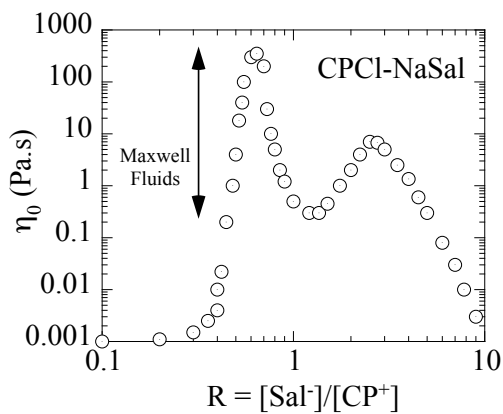

**Figure 7** : Static viscosity $\eta_0$ for cetylpyridinium chloride–sodium salicylate solutions as function of R (= [Sal$^-$]/[CP$^+$]). The CPCl concentration is c = 3.6 wt. % (100 mmol). Data are taken from Rehage and Hoffmann [3]. Solutions with viscosity larger than 0.2 Pas are entangled and exhibit a Maxwellian behavior (Eqs. 8 and 9).

*Maxwellian Behavior* : A step further in the description of the micellar dynamics was made by the first quantitative measurements of the linear mechanical response of these solutions. The pioneering works in that matter were those of Rehage, Hoffmann [3,126,127], Shikata [63,128-130] and Candau [1,131,132] and their coworkers. Rehage and Hoffmann have used rheology to demonstrate that the micellar growth results in an increase of the fluid viscosity. Fig. 7 displays the static viscosity data for cetylpyridinium chloride at 3.6 wt. % (100 mmol) as function of sodium salicylate content [3,126,127]. The steep increase seen at molar ratio R ~ 0.3 is interpreted as a transition between spherical and wormlike micelles. Similar viscosity behaviors were observed with CTAB and NaSal [63,64,133]. Note in Fig. 7 the presence of a secondary maximum around R ~ 2.5. It is interesting to outline here that the two maxima in viscosity correspond to molar ratios at which a nematic phase is observed at higher concentrations (Fig. 6b). The correlation between a high viscosity in the semidilute regime (indicating the formation of very long micelles) and the presence of a nematic phase was recognized by Göbel and Hilltrop [71].

The most fascinating result that Rehage, Hoffmann, Shikata and Candau and their coworkers have discovered by quantitative measurements was that the viscoelasticity of these surfactant solutions was characterized by a single exponential response function. The stress relaxation function G(t) was found of the form :

$$G(t) = G_0 \exp\left(-t/\tau_R\right) \tag{8}$$

where $G_0$ denotes the elastic modulus extrapolated as $t \rightarrow 0$ (equivalent to the storage modulus at infinite frequency $G'_\infty$) and $\tau_R$ is the relaxation time. Eq. 8 describes the behavior of a Maxwell fluid for which the static viscosity $\eta_0$ is the product of $G_0\tau_R$. Rheological experiments are classically performed as function of the angular frequency, and the response function is then the dynamical elastic modulus $G^*(\omega)$. $G^*(\omega) = G'(\omega)+iG''(\omega)$ is the Fourier transform of G(t). $G'(\omega)$ (resp. $G''(\omega)$) denotes the storage (resp. loss) modulus :

$$G'(\omega) = G_0 \frac{\omega^2\tau_R^2}{1+\omega^2\tau_R^2} \quad \text{and} \quad G''(\omega) = G_0 \frac{\omega\tau_R}{1+\omega^2\tau_R^2} \tag{9}$$

Eqs. 8 and 9 have been found repeatedly in viscoelastic micellar systems. This rule is indeed so general that it is now commonly admitted that a Maxwellian behavior is a strong indication of the wormlike character of self-assembled structures. Fig. 8 shows one example of Maxwellian behavior obtained for the CPCl-NaSal system at different temperatures [123]. The storage and loss moduli are displayed in reduced units ($G'/G_0$ and $G''/G_0$ *versus* $\omega\tau_R$) and superposition is obtained on the whole temperature range. The continuous lines are calculated according to Eq. 9. Typical C16 wormlike micellar solutions such as the ones based on CPCl or CTAB surfactants have elastic moduli in the range $1 - 1000$ Pas and relaxation time $\tau_R$ comprised between 1 ms (the lowest relaxation time detectable with rotational rheometers) and several seconds at room temperature. For longer hydrophobic chains, such as for the C22 mono unsaturated cationic surfactant recently studied, Maxwellian behaviors with relaxation times as long as 1000 seconds have been reported at room temperature [72,134,135].

On the theoretical side, the challenge was to account for the unique relaxation time of the mechanical response. This was done by Cates and coworkers in the 1980's with the reptation-reaction kinetics model [1,23,136-139]. The reptation-reaction kinetics model is based on the assumption that in the viscoelastic regime, wormlike micelles form an entangled network analogous to that of polymers. The micelle-polymer structural analogy was demonstrated by means of static and dynamic light scattering during the same period [20,24,34,35,140,141]. Cates suggested that the breaking and recombination events of the chains are coupled to the reptation [104] and accelerate the overall relaxation of the stress [23,136]. In the fast breaking limit, a given micelle undergoes several scission and recombination reactions on the time scale of the reptation. Thus, all initial deformations of the tube segments relax at the same rate, this rate being driven by the reversible scission. Because of the breaking and recombination dynamics, wormlike micelles are often described as equilibrium or living polymers. Other relaxation mechanisms were also explored, such as bond and end interchange processes between micelles [139]. In the reptation-reaction kinetics model, the rheological time $\tau_R$ reads :

$$\tau_R \sim \sqrt{\tau_{break}\tau_{rept}} \qquad (10)$$

where $\tau_{break}$ and $\tau_{rept}$ are the characteristic times for breaking/recombination and reptation, respectively. In wormlike micellar systems, typical breaking times are of the order of milliseconds [24] and Eq. 10 was found to be in qualitative agreement with experimental data [142]. Experimentally however, it has been difficult to determine $\tau_{break}$ and $\tau_{rept}$ separately and to confirm the validity of Eq. 10. Another way to test the predictions of the reptation-reaction kinetics model was to study the scaling properties (as function of the concentration) for some structural and rheological quantities describing the micellar entangled state.

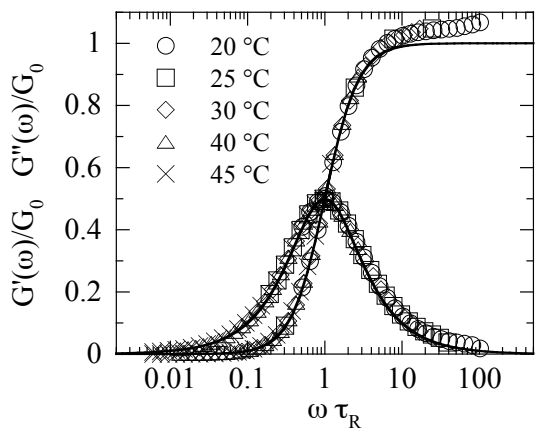

**Figure 8** : Elastic moduli G' and G'' as function of the angular frequency for temperatures comprised between 20 °C and 45 °C. G' and G'' are divided by the elastic modulus $G_0$ and the angular frequency ω is normalized by the relaxation time $\tau_R$ of the fluid. Data are for the CPCl-NaSal wormlike micelles in water (0.5 M NaCl) at c = 12 wt. % [123]. The solid lines correspond to the Maxwellian viscoelastic behavior (Eq. 9).

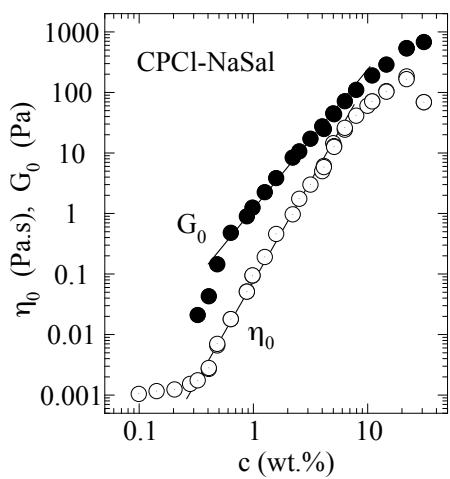

**Figure 9** : Concentration dependences of the static viscosity $\eta_0$ and of the elastic modulus $G_0$ obtained for the CPCl-NaSal system. In the semidilute regime (c* ~ 0.3 wt. %), both quantities exhibit scaling laws as function of the concentration [143]. The exponents are here 2.2 and 3.3, in good agreement with the predictions (see Table II).

*Scaling Laws* : In the entangled regime, the properties of wormlike micelles depend on the micellar volume fraction, *i.e.* the surfactant concentration, and not on the properties of individual micelles such as their average length or their length distribution. By analogy with polymers, Cates suggested that wormlike micelles should follow scaling laws [1,23]. These scaling behaviors have the form of power laws, with scaling exponents that can be compared to those found in experiments. The scaling exponents of some representative quantities related to the structure and rheology of wormlike micelles are given in Table II. These quantities are the elastic modulus $G_0$, the relaxation time $\tau_R$, the static viscosity $\eta_0$, the mesh size $\xi$, the Rayleigh ratio extrapolated at zero wave-vector $\mathcal{R}(q\rightarrow 0)$ and the self-diffusion coefficient $D_S$. Here, $\mathcal{R}(q\rightarrow 0)$ is determined from static light scattering experiments and $D_S$ is the diffusion coefficient of the surfactant molecules. The theoretical exponents are given within the mean-field Gaussian approximation. The reptation-reaction kinetics model predicts for instance that the static viscosity above c* scales as $\eta_0 \sim c^{7/2}$ and the elastic modulus as $G_0 \sim c^{9/4}$. Fig. 9 illustrates such behaviors for the system CPCl-NaSal in 0.5 M NaCl brine [122,143]. In this figure, power laws are observed over more than a decade in concentration and the exponents are close to the predicted ones.

In Table II we also summarize the results of a survey over 15 different surfactant systems and provide the experimental exponents for neutral micelles as a comparison. The elastic modulus $G_0$ is found in agreement with the theory for most (if not all) systems. Scaling exponents are between 1.8 and 2.4. As far as the structure of the network is concerned and although fewer measurements were done, the mesh size $\xi$ and the Rayleigh ratio $\mathcal{R}(q\rightarrow0)$ also agree with theory. Best examples are CTAC [1,20,34,35,65] and CPClO$_3$ [141,144]. The exponent associated to the self-diffusion constant has a broader range of variation, between −1 and −4, whereas the theory predicts −1.58 [145]. In contrast, for the viscosity and for the relaxation time, the predicted exponents are observed in few systems only, as for instance CPCl-NaSal in brine (0.5M NaCl) [121,143], CTAB with 0.25 M KBr [1], CTAB-NaSal in brine (0.1M NaCl) [1], lecithin reverse micelles [87] and metallic salts in organic solvents [89]. For the majority of wormlike micelles, however, the scaling exponents for $\eta_0$ and $\tau_R$ are not in agreement with the model. As function of the concentration, $\tau_R$ can either increase or decrease, or even pass through a maximum [1,3,6,84,126,127,131,134,146]. Some authors have ascribed these discrepancies to the existence of micellar branching [5].

| | scaling exponents (theory) | scaling exponents (experiment) | Refs. |
|---|---|---|---|
| Elastic modulus $G_0$ | + 2.25 | 1.8 … 2.4 | *a* |
| Relaxation time $\tau_R$ | + 1.25 | > 0 and/or < 0 | *a* |
| Static viscosity $\eta_0$ | + 3.5 | 1 … 4 | *a* |
| Mesh size $\xi$ | − 0.75 | − (0.5 … 1) | *b* |
| Rayleigh ratio $\mathcal{R}(q\rightarrow0)$ | − 0.25 | − (0.3 … 0.4) | *b* |
| Self-diffusion constant | − 1.58 | − (1 … 4) | *c* |

**Table II** : Experimental and theoretical exponents of some relevant quantities in semidilute regime of wormlike micelles. The letters in the last column denote sets of references detailed below.
*(a)* : [1,6,84,85,87,89,122,126,127,131,132,134,143,146-149];
*(b)* : [1,20,34,35,65,87,134,141,143,144,150]
*(c)* : [1,140,144,151,152].

We propose here an alternative explanation. In the reptation-reaction kinetics model, the reptation and the breaking times both depend on the average micellar length $\overline{L}$. The scaling are : $\tau_{break} \sim \overline{L}^{-1}c^0$ and $\tau_{rept} \sim \overline{L}^3 c^{3/2}$ [23,136]. So, both $\tau_R$ and $\eta_0$ will depend on $\overline{L}$. This is not the case for $\xi$ and $G_0$. We suggest that the failure to observe the correct scaling for the viscosity and relaxation time is due to the fact that for these systems, the average micellar length does not follow the predicted $c^{1/2}$-dependence (Eq. 1). We suggest also that the theory developed in Section 1.1 for neutral and polyelectrolyte micelles does not strictly apply to the surfactant systems that have been investigated the most. For systems with strongly binding counterions such as CPCl-NaSal and CTAB-NaSal (class **C**) or for cationic and anionic mixtures (class **F**), the picture of complexation between oppositely charged species might be more appropriate to describe the growth law of micelles.

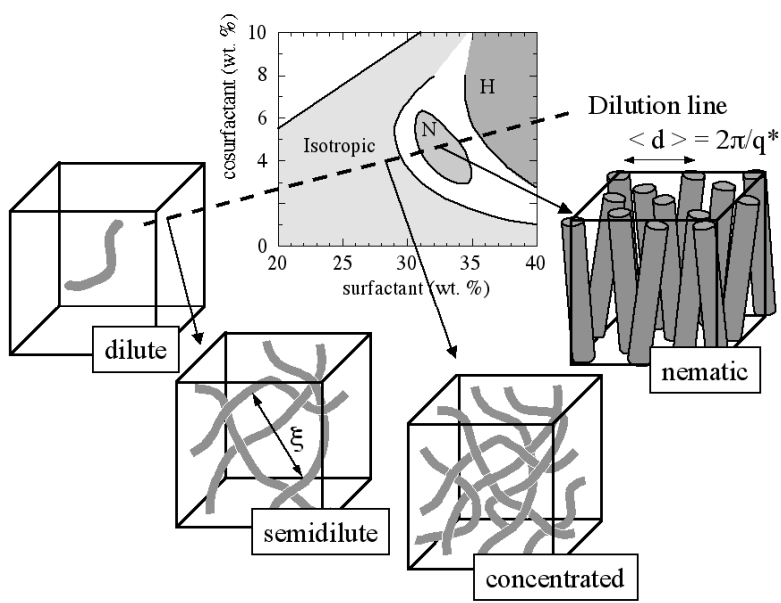

**Figure 10** : Schematic illustrations of the different phases encountered in wormlike micellar solutions, from dilute solutions to concentrated mesophases. The phase diagram shown is that of CPCl-Hex (Fig. 6a). <d> denotes the average distance between colinear micelles in the concentrated isotropic and nematic phases. ξ is the mesh size of the entangled network in the semidilute regime.

## 1.6    Concluding Remarks on the Equilibrium Properties

In this first part, three main aspects of the equilibrium properties of wormlike micelles have been addressed : the self-assembling properties and the dynamics of growth of the linear aggregates, the role of the flexibility on the phase behavior and the relaxation dynamic in the viscoelastic regime. In the semidilute regime, wormlike micelles entangle and form a transient network. The solutions appear as "gels" but quantitative measurements reveal a Maxwellian behavior for most systems. The relaxation times are comprised between 1 ms – 1000 s, depending on the surfactant chain length. We have also noted that the Maxwellian behaviors extend well beyond the surfactant phases reviewed here. They have been observed on a wide variety of low molecular weight molecules that are known to self-assemble, as surfactants, into linear, semiflexible and reversible aggregates.

In the concentrated regime, the mesh size of the network is shorter than the persistence length b and orientational correlations start to appear. We have cited several examples of surfactant systems, including systems with screened and unscreened electrostatics, for which an increase in concentration yields a succession of thermodynamically stable mesophases such as nematic and hexagonal. This sequence of phases is illustrated in Fig. 10, together with a representation of the local environments for the aggregates. This picture is in relative good agreement with the theories that describe the phase behavior of linear and semiflexible aggregates [26,114,136]. These two features, the Maxwellian viscoelasticity and the presence of long-range orientational order at high concentration are playing a crucial role in the nonlinear rheology of wormlike micelles and in the shear banding transition.

## 2.    SHEAR BANDING TRANSITION IN CONCENTRATED AND SEMIDILUTE REGIMES

When submitted to a steady shear, wormlike micelles generally do not change their local morphology. The aggregates remain cylindrical, with an average length and distribution that might eventually depend on the shear rate [153,154]. A number of publications suggest that the solutions undergo a transition of shear banding. The shear banding transition is a transition between an homogeneous and a non homogeneous state of flow, the latter being characterized by a demixion of the fluid into macroscopic regions of high and low shear rates. Shear banding has been identified unambiguously in wormlike micelles using flow birefringence [14] and NMR velocimetry [15,16] measurements in the years 1990's. Since the first reports on micellar fluids, non-homogeneous flows were observed in many other complex fluids, such as cubic phases of soft colloids [155], electro- and magneto-rheological fluids [156], transient networks of associative polymers [157] and soft glassy materials [158]. Shear banding is in general an abrupt transition that occurs at a critical shear rate, noted $\dot{\gamma}_1$ in the following. In some cases, it is appropriate to introduce a second characteristic shear rate noted $\dot{\gamma}_2$, which corresponds to the onset of a second homogeneous flow at higher shear rates. In the following, we review the phenomenology of shear banding in concentrated and semidilute wormlike micelles. The first step of this survey consists in defining its rheological signatures.

## 2.1    Isotropic-to-Nematic Transition in the Concentrated Regime

### 2.1.1    Steady State and Transient Rheology

By analogy with polymer solutions, the concentrated regime of wormlike micelles is defined for entangled networks with mesh sizes of the order or shorter than the persistence length, *i.e.* $\xi < b$ [143]. In the surfactant systems cited in the first part, this inequality corresponds to weight concentrations comprised between ~ 10 wt. % and $c_{I-N}$, the isotropic-to-nematic phase boundary. In this section, we illustrate the rheology of concentrated wormlike micelles by focusing on a solution made from cetylpyridinium chloride/hexanol. This CPCl-Hex solution is isotropic but close to the isotropic-to-nematic boundary (Fig. 10). The steady state shear stress shown in Fig. 11 as a function of the shear rate $\dot{\gamma}$ has been obtained by conventional rheometry using a cone-and-plate geometry. The flow curve of the CPCl-Hex solution exhibits a discontinuity of slope at the critical value $\dot{\gamma}_1$, followed by a stress plateau that stretches over more than a decade in shear rates. We call $\sigma_P$ the value of the stress at the discontinuity. At high shear rates, there is a further increase of the stress. Various shear histories have been applied to the solution in order to ensure the unicity and robustness of the plateau. The data in Fig. 11 can thus be considered as steady state values.

In some concentrated wormlike systems, the plateau is not totally flat, but exhibit a slight increase as function of the shear rate. This increase is described in general by a power law, $\sigma(\dot{\gamma}) \sim \dot{\gamma}^{\alpha}$ where $\alpha = 0.1 - 0.3$.

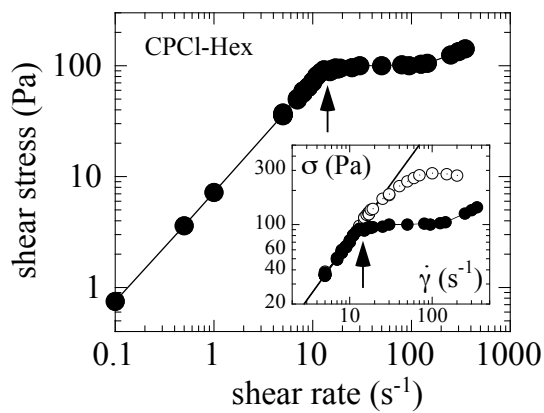

**Figure 11** : Steady state response of a concentrated cetylpyridinium chloride / hexanol (CPCl-Hex) micellar solution. Concentrations are $c_{CPCl}$ = 28.0 wt. % and $c_{Hex}$ = 3.9 wt. %. Experiments are made using a rotational rheometer using a cone-and-plate geometry. The onset of the stress plateau is at the critical shear rate $\dot{\gamma}_1$ = 14 s$^{-1}$. The static viscosity of this micellar fluid is 7.2 Pas [148,159]. Inset : Zoom of the plateau region, showing the initial (open symbols) and the steady state (close symbols) values of the stress, as obtained from start up experiments.

Stress plateaus or pseudo-plateaus in concentrated wormlike micelles were reported for the first time in CPClO₃ [116] and CPCl-Hex [121,159] systems. They have been found for other surfactants, such as CPCl-NaSal [123,148], CTAB [117,124,160], CTAB-KBr [161] and CTAT [162]. Stress plateaus are the central feature in the nonlinear rheology of wormlike micelles.

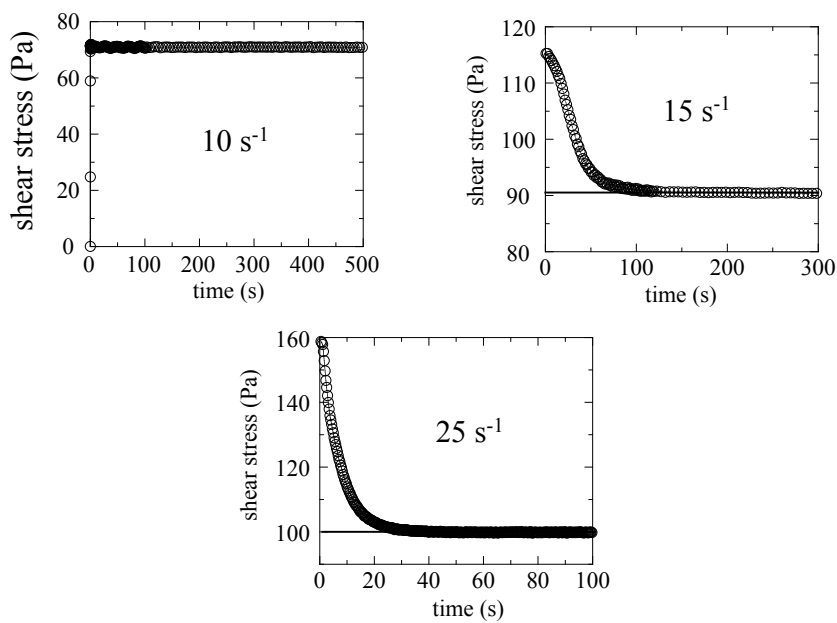

**Figure 12** : Transient shear stress as determined from startup measurements for the solution shown in Fig. 11. Data are taken below and in the plateau region. The characteristic time of the initial stress decrease at 15 and 25 s⁻¹ is much larger than the Maxwell relaxation ($\tau_R$ = 0.02 s). The slow kinetics corresponds to the formation of shear bands.

In Figs. 12 are displayed the transient stress responses at different shear rates, below the critical rate $\dot{\gamma}_1$ (12a) and in the plateau region (12b and 12c). These transients were obtained using startup experiments. Below $\dot{\gamma}_1$, the stress grows rapidly (in times < 1 s) up to a stationary value and remains constant. Above $\dot{\gamma}_1$, the transient stress exhibits the same rapid increase at short times and a slow relaxation at longer times. The stress kinetics in the plateau region is described by an expression of the form [122,148,163] :

$$\sigma(t) = \sigma_{ST} + \left(\sigma(t \to 0) - \sigma_{ST}\right)\exp\left[-\left(t/\tau_N\right)^n\right] \qquad (11)$$

where the initial and steady state stresses $\sigma(t \to 0)$ and $\sigma_{ST}$, as well as the characteristic time of the relaxation $\tau_N$ depend on $\dot{\gamma}$. For flat plateaus $\sigma_{ST}$ = $\sigma_P$. In the inset of Fig. 11, $\sigma(t \to 0)$ and $\sigma_{ST}$ are shown by open and closed symbols. Note that the stress overshoot $\sigma(t \to 0) - \sigma_{ST}$ at its maximum is much larger than the stationary value of the stress itself, suggesting a rather profound transformation of the fluid in this regime. The slow kinetics of the shear stress in the plateau region was originally assumed to coincide with the nucleation and growth of shear bands [122,148]. The coefficient n in Eq. 11 has been found to vary in the range $1 - 3$, depending on the system and on the shear rate [164]. Note finally that in the systems showing stress plateaus and slow kinetics, the characteristic time $\tau_N$ in Eq. 11 is always much larger than the Maxwell relaxation time $\tau_R$ derived from the linear response. For the CPCl-Hex solution studied here, $\tau_R = 20$ ms whereas $\tau_N$ varies between 1 and 100 s depending on the shear rates [164].

The data presented in Figs. 11 and 12 were obtained in controlled strain rheometry, *i.e.* in an experiment where the strain rate is held fixed and where the stress is deduced from the torque transmitted by the fluid. Rotational rheometry also allows to control the stress and to simultaneously record the angular velocity of the mobile part. Semidilute and concentrated wormlike micelles have been extensively studied using controlled stress rheometry [14,42,84,117,132,134,161,165,166]. The generic behavior for this type of experiment is summarized in Fig. 13. For a CTAB-D$_2$O concentrated solution (c = 18 wt. %, T = 32 °C), the shear stress measured at steady state exhibits two stable branches at low and high shear rates. These two branches are separated by a stress plateau which coincides exactly with that of a strain controlled experiment.

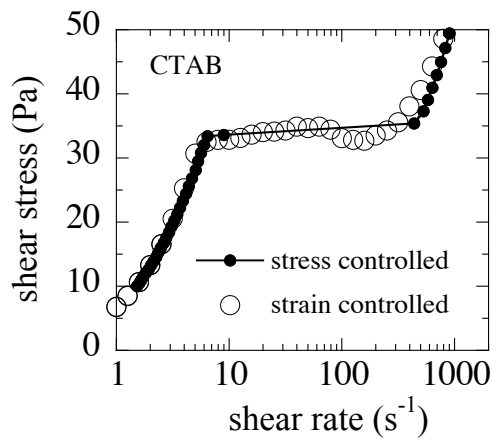

**Figure 13** : Comparison between the steady state shear stress results obtained from controlled strain and controlled stress rheometry in the plateau region. The data are for CTAB-D$_2$O wormlike micelles at c = 18 wt. % and T = 32 °C [124]. When stress is applied, the sheared solution jumps from the low shear to the high shear branch of the flow curve. In the lower branch, the stress increases linearly with a static viscosity $\eta_0 = 6.7$ Pas.

The interesting feature here is that no steady state can be set up in the plateau region when the stress is controlled. Instead, there is a jump in shear rate, the shear rate switching rather abruptly from $\dot{\gamma}_1$ to $\dot{\gamma}_2$. The present results reinforce the idea of a shear-induced phase separation, where the stress (resp. strain rate) is the intensive (resp. extensive) variable.

### 2.1.2    Flow birefringence

Flow birefringence experiments have been carried out during the past decades to investigate the orientation properties of polymer solutions and melts [167,168]. More recently, rheo-optical techniques were applied with success to the study of wormlike micelles [2,14,117,161,169-172]. Experimental set-ups designed for studying complex fluids generally utilize a Couette cell as shearing device (Fig. 14). For flow birefringence, the cell has a narrow gap and the polarized light is sent along the vorticity direction. The velocity (**v**), velocity gradient ($\vec{\nabla}$**v**) and vorticity ($\omega$ =**v**×$\vec{\nabla}$**v**) directions are indicated in Fig. 14 for the Couette geometry. The two physical quantities determined in this type of experiments are the birefringence intensity $\Delta$n and the extinction angle $\chi$. $\Delta$n is the difference between the indices of the ordinary and extraordinary directions and $\chi$ is the angle of the refractive index tensor with respect to the flow velocity. For an isotropic fluid, $\Delta$n = 0 and $\chi$ = 45 °.

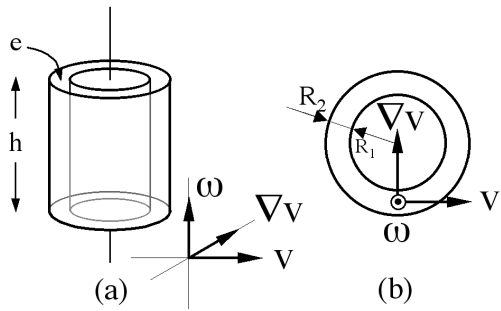

**Figure 14** : Representation of a Couette cell used in the investigations of the structure of wormlike micelles under shear. The fluid is introduced between the two concentric cylinders (a). The value of the gap e = $R_2 − R_1$ is small compared to the overall size of the cell. The motion of the outer cylinder (the inner one being immobile) produces a velocity gradient in the gap. The three directions characterizing simple shear are shown : the velocity **v**, the velocity gradient $\vec{\nabla}$**v** and the vorticity $\omega$. Data from flow birefringence, small-angle neutron scattering and nuclear magnetic resonance were obtained with this type of device. In flow birefringence, the incident and polarized light beam propagates along the vorticity direction. In SANS the incident beam passes through the Couette along $\vec{\nabla}$**v** (radial configuration). It goes twice through the fluid in motion.

The device in Fig. 14 can also be used for direct visualization of the solution under shear. The transmitted light is recorded on a digital camera with a spatial resolution of the order of microns. This was the configuration that was chosen to show the phenomenon of shear banding transition in wormlike micelles [14,161,173]. Fig. 15 shows six photographs of a 1mm-gap containing a concentrated CPCl-Hex solution under steady shear, before the onset of stress plateau (a), in the plateau region (b - e) and after the upturn at high shear rate (f). We note that in Fig. 15a and 15f, corresponding to the lower and upper branches of flow curve, the birefringence and the micellar orientations are homogeneous throughout the gap. For intermediate $\dot{\gamma}$, there is a coexistence between a bright and a dark band. The dark and bright bands in Figs. 15 have birefringences $\Delta n$ of the order of $-10^{-5}$ and $-10^{-3}$, respectively. The negative sign of $\Delta n$ arises from the anisotropy of the polarisability tensor that describes a monomer in the micellar chain [170,171]. The dark band in Figs. 15a–15e results from the adjustment between the angle made by the polarizer and $\chi$. In the plateau regime, the bright band broadens as the shear rate increases. The growth of the strongly birefringent phase from the inner cylinder is generally explained in terms of curvature effect of the Couette cell. For concentric cylinders, the shear rate is a weakly varying function of the spatial coordinate and its largest value is reached at the inner wall [174,175]. The shear banding is initiated where the fluid first reaches $\sigma_P$, at the inner wall.

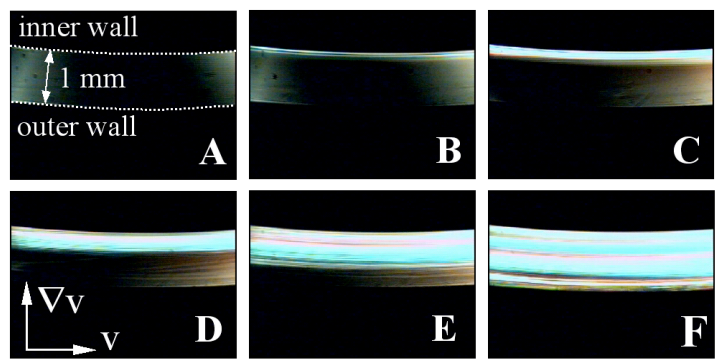

**Figure 15** : The photographs show the gap of a Couette cell containing a CPCl-Hex solution at concentrations $c_{CPCl}$ = 28.0 wt. % and $c_{Hex}$ = 3.9 wt. % (see Figs. 11 and 12 for the rheology). The labels correspond to increasing shear rates. Photographs (a) and (f) are taken in the Newtonian regime and in the high shear rate branch, respectively. Photographs (b) to (e) are taken in the plateau region. Typical exposure times are of the order of milliseconds. The polarizer and analyzer are oriented so that the band close to the outer cylinder appears dark. With increasing shear rates, the bright nematic band fills up the gap progressively.

In order to prove that the shear-induced bright band corresponds to a nematic phase, we compare in Fig. 16 the flow birefringence of an isotropic solution at high shear rate and a nematic solution under moderate steady shear. The two fluids are made from the same micelles and the nematic liquid crystal is slightly more concentrated than the isotropic one. Fig. 16 shows that the transmitted light, and thus the birefringence Δn are the same for the two experiments. The extinction angles in both solutions are also close to 0 °, as expected for an oriented nematic. The textures observed for the nematic solution vanish if $\dot{\gamma}$ is further increased.

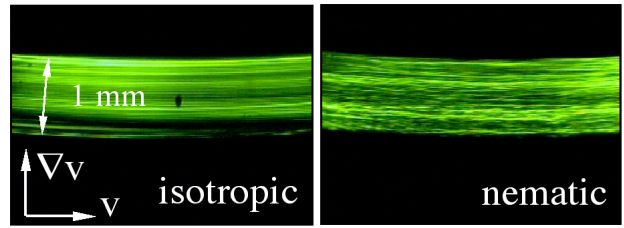

**Figure 16** : Comparison between the flow birefringence of two CPCl-Hex solutions under shear. The solution on the left is isotropic at rest (same concentrations as in Fig. 15) and the photograph is taken at 220 s$^{-1}$. This corresponds to the high shear rate branch of the flow curve. The solution on the right is nematic at rest ($c_{CPCl}$ = 29.1 wt. % and $c_{Hex}$ = 4.1 wt. %) and the photograph represents the fluid sheared at 8.9 s$^{-1}$. In order to reduce the transmitted intensity, a green filter has been placed on the optical path before the solution.

### 2.1.3   Neutron Scattering under Shear

As discussed in the first part, SANS is generally used to probe the translational degrees of freedom of microstructures on a 10 Å – 1000 Å length scale. Using a two-dimensional detector, SANS can also probe the orientational degrees of freedom of a micellar fluid subjected to a flow. In a typical SANS experiment, a set-up similar to that of flow birefringence is utilized (Fig. 14). In the radial (resp. tangential) scattering configuration, the incident beam passes through the Couette along the gradient (resp. velocity) direction. Figs. 17a–17c show two-dimensional scattering patterns obtained for a CPCl-Hex solution at a concentration close to isotropic-to-nematic transition. The shear rates correspond to the three different branches of the flow curve. In the first stable branch, at low shear rate, the scattering is isotropic and exhibits a broad maximum at q* ∼ 0.1 Å$^{-1}$. This maximum results from strong translational correlations between the micellar threads. An estimate of the distance between micelles (*i.e.* between their axis of symmetry) yields < d > = 2π/q* ∼ 60 Å. In the plateau regime, some anisotropy arises in the scattering function, which manifests itself by symmetric crescent-like peaks in the vorticity direction $\mathbf{q}_\omega$. At still higher shear rate, the ring-like structure vanishes and the scattering is dominated by the anisotropic pattern (Fig. 17c). This latter pattern is qualitatively analogous to the one obtained from a micellar solution which is nematic at rest and subjected to a moderate shearing. The scattering signature of a nematic phase of micelles is shown in Fig. 18 for comparison. Similar results were reported on various surfactant systems close to the I-N transition [116,124,148,159].

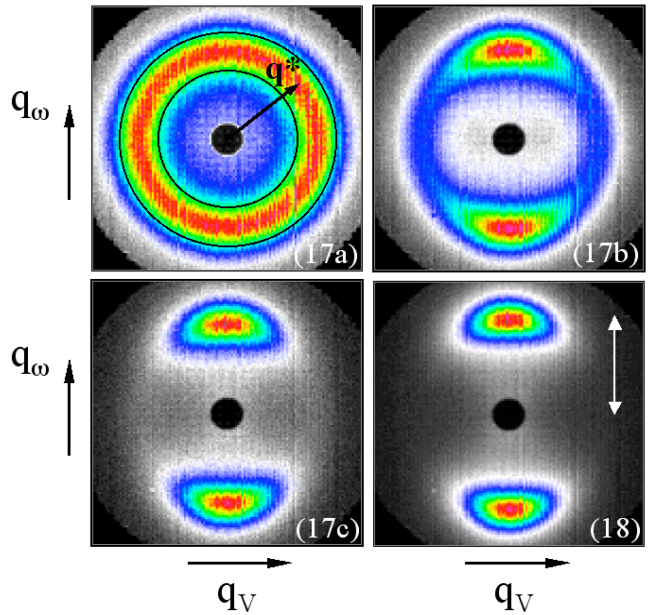

**Figure 17** : Neutron scattering intensities obtained from a concentrated micellar solution under shear [121,159]. Shear rates are $\dot{\gamma} = 0$ (a), 64 (b) and 213 $s^{-1}$ (c), respectively. The data are recorded in the velocity-vorticity plane, corresponding to the wave-vectors $\mathbf{q}_v$ and $\mathbf{q}_\omega$. The maximum scattering at $\dot{q}^* = 0.0996$ Å$^{-1}$ arises from steric interactions between micelles.
**Figure 18** : Neutron scattering intensity obtained from nematic wormlike micelles under shear [119,176] at the shear rate $\dot{\gamma} = 100$ s$^{-1}$. As in Fig. 17, the data are recorded in the velocity-vorticity plane. The scale for Figs. 17 and 18 is given by the arrow which represents 0.1 Å$^{-1}$.

Neutron data have been analyzed quantitatively in terms of orientational degrees of freedom. The proportions of isotropic and nematic phases under shear, as well as the order parameter of the shear-induced nematic can thus be derived. The two-dimensional spectra of Figs. 17 and 18 are first converted into vectors representing the angular distribution of the scattered intensity $I(\psi)$, where $\psi$ is the azimuthal angle. $\psi$ is defined as the angle between the wave-vector $\mathbf{q}$ and the vorticity direction $\mathbf{q}_\omega$. For isotropic patterns, $I(\psi)$ is constant whereas for anisotropic patterns it is maximum at $\psi = 0$ and $\psi = 180$°. The azimuthal intensity is then fitted using the phenomenological relation :

$$I(\psi, \dot{\gamma}) = I_I(\dot{\gamma}) + I_N(\dot{\gamma}) \sinh(m \cos^2 \psi) \qquad (12)$$

where $I_I(\dot{\gamma})$ is the contribution of the isotropic phase and $I_N(\dot{\gamma})$ the prefactor to the nematic component. The $\psi$-dependence of the nematic contribution is periodic and it is characterized by a unique parameter $m$. This parameter determines the order parameter of the oriented phase. The knowledge of $I_I(\dot{\gamma})$ and $I_N(\dot{\gamma})$ on the other hand allows us to derive the proportion of each phase at different shear rates. These results are discussed in Section 2.1.4 and compared with data from flow birefringence and nuclear magnetic resonance.

The shear-induced phase is characterized by the orientational distribution of the micellar threads, or equivalently by the moments of this distribution. $P_2$, the second moment of the distribution represents the order parameter of the phase.

$$P_2 = \left\langle \left(3\cos^2\beta - 1\right)/2 \right\rangle, \tag{13}$$

where the average $<...>$ is performed over all micellar orientations and where $\beta$ is the angle made by the cylindrical micelles with the flow velocity. Using analytical expressions to relate the $n^{th}$-order moment of the distribution to the experimental intensity $I(\psi, \dot{\gamma})$ [177], we found for $P_2(\dot{\gamma})$ a constant value in the plateau region, $P_2 = 0.65 \pm 0.03$ [159]. This value agrees well with that of a nematic monodomain made from the same micelles, $P_2 = 0.70 \pm 0.04$ [118,176] and is slightly higher than the order parameter predicted for persistent chains, $P_2 = 0.49$ [114]. It can be concluded that the shear-induced phase is strongly oriented and nematic. Due to its excellent alignment, director instabilities within the new phase, such as tumbling or wagging can be ruled out [118,178].

### 2.1.4    Nuclear Magnetic Resonance under Shear

As for the wormlike micelles, Nuclear Magnetic Resonance (NMR) was used primarily to determine the velocity field in the banded regime [15,16,160,179-184]. With NMR, it is also possible to resolve spatially (*i.e.* within the gap of the shearing cell) the spectral splitting associated with the quadrupole interactions of the deuteron nucleus with the local electric-field gradient. This splitting is actually proportional to the order parameter of the phase that is initiated. Should this splitting be zero, the phase is disordered, should it be non zero, the phase is nematic. The splitting is actually due to the fact that in an oriented nematic of micelles, the $D_2O$ molecules of the solvent inherit of the alignment of the cylindrical structures [50]. The NMR technique allows to measure velocity fields and spectral splitting with a spatial resolution of the order of $10 - 50$ $\mu$m and needs in general long acquiring times. This technique was applied to the CTAB-$D_2O$ solution (**c** = 20 wt. %, T = 41 °C), a solution that compares well with the one of Fig. 13. The shear banding transition observed by neutron and flow birefringence was confirmed by NMR under shear. Furthermore, the proportions of isotropic and nematic phases in the plateau regime were calculated [160] and they were found in remarkable agreement with the SANS results (Fig. 19). NMR Velocimetry profiles performed simultaneously on the same solution are showing also that the birefringent band visualized by rheo-optics does not correspond necessarily to a high shear band. This intriguing result was interpreted as arising from a nematic phase of high viscosity, possibly associated with mesoscale ordering. Since the first results on micelles, NMR has been applied to other complex fluids for velocity field imaging [158].

A list of concentrated wormlike micelles showing stress plateaus and shear banding have been established from recent literature data in Table III. The techniques used to evidence shear banding (flow birefringence, SANS and NMR) and their related references are also provided. The conclusions of this survey will be drawn at the end of Section 2.2.

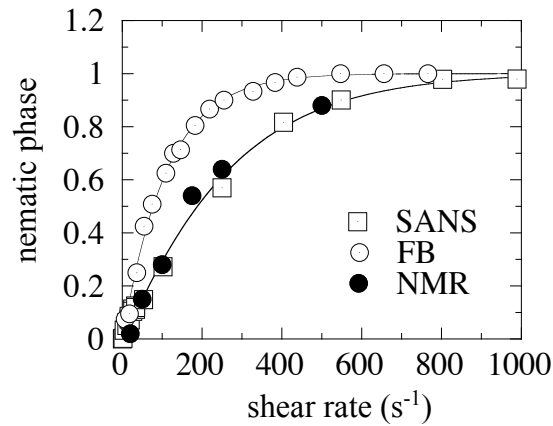

**Figure 19** : Percentage of the nematic phase induced by shearing obtained from flow birefringence [124], small-angle neutron scattering [124] and deuterium NMR [160]. Experiments are performed on a CTAB-$D_2O$ concentrated solution closed to I-N transition. The data from SANS and NMR are in good agreement.

## 2.2    Shear Banding in Semidilute Regime

In the semidilute wormlike micelles, *i.e.* for concentrations comprised between 0.1 wt. % to 10 wt. % (or equivalently 0.003 mol/l to 0.3 mol/l for a C16 surfactant), the structural and rheological properties are qualitatively analogous to those found for concentrated solutions. Stress plateaus and shear banding are still the main features of these solutions in response to shear flow. There are differences however with the concentrated regime. These differences are threefold :
i) Contrary to the concentrated regime, there is not a single behavior in the rheology of semidilute solutions. The nonlinear mechanical response can be either shear-thinning or shear-thickening. In the shear-thinning class, the flow curves and shear banding characteristics can differ from one system to the other.

*ii)* Because of this broad diversity in rheological behaviors, experimental data in the semidilute regime are often incomplete. Few systems have been investigated using several different techniques, as it is the case for concentrated solutions.

*iii)* The last point concerns electrostatics. Diluting micellar phases (and keeping the solutions free of additional salt) increases the range of the electrostatic interaction between micelles. This may have considerable effects on the stability and equilibrium properties of the phases, as it was recently pointed out [185,186]. In the concentrated regime on the contrary, the ionic strength is always large (of the order of 1 M) and electrostatics has a minor effect. In this section, emphasis will be put on shear-thinning systems with screened electrostatic interactions.

| surfactant | additive | salt | conc. regime | shear banding | Refs. |
|---|---|---|---|---|---|
| CPCl | NaSal | | sd | NMR | *a* |
| CTAB | NaSal | | sd | .. | *b* |
| CTAB | | KBr | sd/c | FB | *c* |
| CPClO$_3$ | | NaClO$_3$ | c | SANS | *d* |
| CPCl | Hexanol | NaCl | c | SANS, FB | *e* |
| CPCl | NaSal | NaCl | sd/c | FB,DLS | *f* |
| CTAC | NaSal | | sd | FB | *g* |
| CTAB | | | c | SANS,FB,NMR | *h* |
| CTAHNC | | | sd | .. | *i* |
| CTAB | | NaNO$_3$ | sd/c | FB | *j* |
| CTAT | | | sd/c | .. | *k* |
| CTAT | SDBS | | sd | .. | *l* |
| SDES | | AgCl$_3$ | sd | .. | *m* |
| EHAC | NaSal | | sd | .. | *n* |

**Table III** : Wormlike micellar systems known to exhibit stress plateaus or quasi plateaus in steady state rheology. Systems for which the shear banding has been observed by structural methods are indicated. The abbreviations are "sd" for semidilute and "c" for concentrated. As for the techniques used to show shear banding, "SANS" stands for small-angle neutron scattering, "FB" for flow birefringence, "NMR" for nuclear magnetic resonance and "DLS" for dynamic light scattering. The letters in the last column denote sets of references detailed below.
*(a)* : [3,15,16,127,181,182,187-189]; *(b)* : [64,132,190]; *(c)* : [132,161,191,192];
*(d)* : [116]; *(e)* : [122,159,164]; *(f)* : [122,123,148,163,183,189,193]; *(g)* : [14,172];
*(h)* : [117,124,160,173]; *(i)* : [194,195]; *(j)* : [165,196-198];
*(k)* : [146,199-203]; *(l)* : [84]; *(m)* : [42]; *(n)* : [134]

### 2.2.1    Generalized « Flow Phase Diagram » and Standard Behavior

Cetylpyridinium chloride with sodium salicylate is certainly the system for which the most extensive set of rheological data in the nonlinear regime is available [15,60,122,123,126,127,143,163,183,188,189,204]. To our knowledge, it is also the first system for which a stress plateau has been reported [3]. When electrostatics is screened by addition of sodium chloride, CPCl-Sal has been shown to exhibit the right scaling exponent for the concentration dependence of the viscosity (Fig. 9) and the micellar fluids prepared between 1 wt. % to 30 wt. % reveal a clear Maxwellian behavior. The steady state shear stress has been measured as function of the shear rate, temperature and concentration and temperature-concentration superposition principle has been used in order to derive the generalized flow phase diagram [123]. The steady shear stress normalized by the elastic modulus $G_0$ is shown in Fig. 20 as function of the normalized shear rate $\dot{\gamma}\tau_R$, for c comprised between 2 wt. % and 21 wt. %. The flow curve at c = 21 wt. % makes the link with the concentrated regime. There, a quasi-plateau is observed and the findings are in line with the data of Figs. 11 and 13. As concentration decreases, stress plateaus are still observed, but the normalized stress and shear rate at which the discontinuity occurs are shifted to larger values.

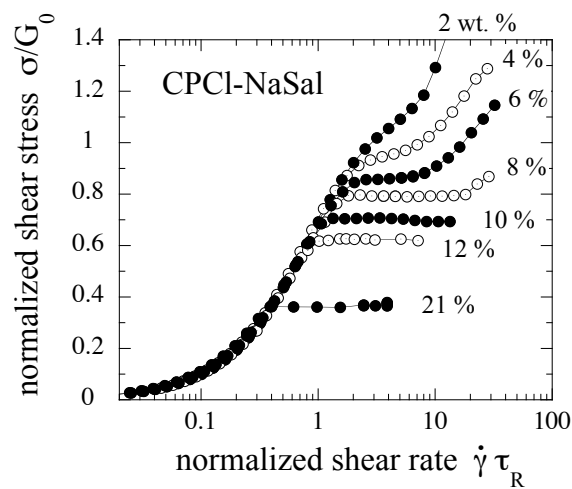

**Figure 20** : Generalized "flow phase diagram" obtained for CPCl-NaSal wormlike micelles. The shear stress normalized by the elastic modulus $\sigma_P/G_0$ is plotted against the reduced shear rate $\dot{\gamma}\tau_R$. The diagram is obtained using the invariance property of the flow curves under temperature and concentration changes. No stress plateau is observed for the critical conditions $\sigma_P/G_0 > 0.9$ and $\dot{\gamma}\tau_R > 3$ [123].

At c = 6 wt. % and above, the transition is much smoother and the stress levels off without discontinuity. According to Fig. 20, stress plateaus are found in semidilute wormlike micelles for stress and rates below the following critical conditions [123] :

$$\sigma_P/G_0 = 0.9 \quad \text{and} \quad \dot{\gamma}\tau_R = 3 \pm 0.5 \quad (14)$$

Assuming the relationship $G_0 \sim k_B T/\xi^3$ between the modulus and the mesh size $\xi$ of the network, $\sigma/G_0$ can be viewed as the mechanical energy dissipated at the scale of the mesh in units of thermal energy. The product $\dot{\gamma}\tau_R$ represents the shear rate in the time scale unit of the fluid. The analogy of Fig. 20 with the phase diagram of an equilibrium system undergoing a phase separation (such as the liquid-gas transition) is striking. This analogy is strengthened by two observations. First, the ratio $\sigma_P/G_0$ decreases linearly with the concentration and extrapolates at the concentration $c_{I-N}$ at which the system at rest undergoes the isotropic-to-nematic transition ($c_{I-N} = 36$ wt. %, Fig. 21). The critical conditions quoted in Eq. 14 also suggest that by choosing the concentration, temperature or salt content adequately, it is possible to find a stress plateau comprised between $\sigma_P/G_0 \sim 0$ and 0.9, and of onset comprised between $\dot{\gamma}\tau_R \sim 0$ and 3. Second, the flow curves with stress plateaus all exhibit the slow transient kinetics and sigmoïdal relaxations described from the concentrated regime. The coefficient n in Eq. 11 was found between 1 and 3 [122,148,188,197]. In CPCl-NaSal, the stretched coefficient n = 2 was interpreted as an indication a one-dimensional nucleation and growth process [121,123,163]. Semidilute wormlike micelles showing stress plateaus similar to those of Fig. 20 has been included to the list in Table III. This collection of systems, all sharing the same rheological properties aims to demonstrate that the behavior described here is actually representative for this class of materials. In the following, Maxwellian linear behaviors and shear banding transition associated with nonlinear stress plateaus will define the "standard behavior" in wormlike micelles.

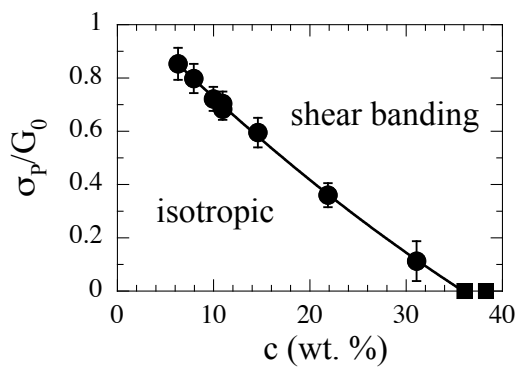

**Figure 21** : Concentration dependence of the reduced stress (closed circles) at the plateau for the CPCl-NaSal system [122]. At high concentrations, the values of $\sigma_P/G_0$ extrapolate to $c_{I-N}$, the isotropic-to-nematic concentration (closed square, $c_{I-N} = 0.356$). At concentration below 6 wt. %, stress plateaus are not observed.

### 2.2.2 Structure of the Shear Bands in the Semidilute Regime

*Flow Birefringence* : Decruppe and coworkers have shown the existence of shear bands in several micellar systems, e.g. in CTAC-NaSal [14,172], in CTAB-KBr [161,191,192] and in CTAB-NaNO$_3$ [197,198]. From their works, two main behaviors were recognized. The thorough study of CTAB-KBr shows strongly birefringent shear bands starting at the inner wall of the cell and progressively filling the gap as the shear rate is increased. As for concentrated systems, shear banding was found to correlate with the onset of stress plateau at $\dot{\gamma}_1$. And as for nematics, the bright band was shown to have an extinction angle $\chi \sim 0°$. This system was re-examined recently with respect to the kinetics of band destabilization, of reconstruction and travel of the interface [192]. CTAB-KBr behaves thus as the micellar systems showing the shear-induced isotropic-to-nematic phase transition [161]. For semidilute solutions in general, a complete transformation of the micellar fluid into a nematic phase requires very high shear rates which are not compatible with the experimental setups described so far.

The second type of behaviors was found in CTAB-NaNO$_3$ [197,198] and CPCl-NaSal [123]. In both cases the electrostatic interactions were screened. CTAB-NaNO$_3$ is interesting for several reasons. In the semidilute regime, for a concentration of 10.9 wt. % with [NaNO$_3$] = 1.79 M, the linear and nonlinear rheology, and even the short time flow birefringence are in agreement with the "standard" behavior outlined in the Section 2.2.1. The long time flow birefringence however shows some striking features. As time evolves, the shear band splits into several and fine striated sub-bands of few microns in thickness. These striations can occupy half of the gap at steady state and show unstable fluctuations of positions and width. This type of band structure is clearly different from that of concentrated solutions and therefore may not easily be associated with a shear-induced I-N transition.

*Nuclear magnetic Resonance* : As for the concentrated solutions, NMR has been used as an imaging technique to determine the velocity field in the banded regime. The micellar solution studied was cetylpyridinium – sodium salicylate in pure water, at concentrations 100 mmol and 60 mmol. This solution is semidilute and it is considered by several research groups as representative of the wormlike micelles class [3,188,204]. NMR was performed on this substance using different flow geometries, such as Couette [15,16,205], cone-and-plate [179,180,182] and Poiseuille [15,181,183]. In the three types of geometries cited, shear banding was found. With the Couette cell, a high-velocity and very thin band was detected near the inner cylinder. The cone-and-plate geometry has a more uniform distribution of shear rate than the Couette, and it displayed slightly different results. The high shear rate band now shows up at the center of the sheared solution. Two interfaces with the adjacent low shear regions are visible on the NMR-plots [15,189]. With pipe flow, the increase of the flow rate from the Newtonian to the plateau regime corresponds to the transition between parabolic and almost flat velocity profiles. For the latter, the shear rate is localized close to the walls as in a plug flow. We should finally mentioned recent velocity field measurements using dynamic light scattering in heterodyne mode. The spatial resolution of this new technique is comparable to that of NMR (~ 50 µm). The data obtained on a CPCl-Sal semidilute solution at c = 6 wt. % (its flow curve is shown in Fig. 20) confirm the coexistence of two macroscopic shear bands in the plateau region [193].

In conclusion, the shear banding in the semidilute regime of wormlike micelles shows more complex features than in the concentrated regime. The NMR velocimetry measurements have evidenced the role of the flow geometry on the band structures. We have completed Table III by adding comments and references for every system that display shear banding, observed either by flow birefringence, by NMR or by SANS.

### 2.2.3    Non Standard Behaviors

As already mentioned, not all semidilute systems agree with the "standard behavior". There are few exceptions and it is probably instructive to describe some of them in order to illustrate the diversity of rheological behaviors. The first system of interest is the binary solution of hexadecyloctyldimethylammonium bromide (C18-C8DAB, see Table I) in water at c = 2.3 wt. %. This micellar solution was investigated by Hoffmann and coworkers by SANS at rest and under shear. It was shown that above a threshold shear rate, the solution undergoes an isotropic-to-hexagonal transition. The long range hexagonal order was characterized by a strong anisotropy perpendicular to the velocity direction (as in Fig. 17), with clearly identified a first and a second Bragg peaks. The data were interpreted in terms of coexistence of two types of cylindrical micelles of different lengths, the short ones contributing to the ring-like pattern and the long ones being responsible for the hexagonal phase. Although the data resemble those of the I-N transition in the concentrated regime, to our knowledge neither shear banding nor stress plateaus were reported in C18-C8DAB solutions.

A second system showing a non-standard behavior is the equimolar solution made from cetylpyridinium chloride and sodium salicylate, both at 40 mmol (c = 2.1 wt. %). At this concentration, CPCl-NaSal is Maxwellian and has a relaxation time around 10 ms. With increasing shear rate, the solution first shear-thins, the stress reaching continuously a plateau regime. The solution then undergoes an abrupt and strong shear-thickening. This instability occurs at reduced shear rate $\dot{\gamma}\tau_R$ around 3 and is associated with the appearance of clear and turbid bands stacked perpendicular to the vorticity direction. At steady state, the bands oscillate. This double transition, shear-thinning and then shear-thickening, as well as the simultaneous occurrence of bands along the vorticity is a unique feature in wormlike micelles. This instability has received much attention during the last years [206].

Wunderlich *et al.* and Pine and coworkers have also studied the flow behaviors of equimolar solutions of CTAB-NaSal [207,208] and of TTAA-

NaSal solutions (TTAA stands for tris(2-hydroxyethyl)-tallowalkyl ammonium acetate, [209-212]). The weight concentrations (comprised between 0.1 and 1 wt. %) are here slightly lower as compared to the previous case. Wunderlich and Brunn had first noticed the existence of anomalously long transients at the thickening transition, as well as a striking dependence of the viscosity as function of the gap of their Couette cells. At concentrations between $1 - 10$ mmol, the micellar network is hardly entangled and the static viscosity is larger than that of the solvent, only by a factor $10 - 100$. As shear is increased, there is first a modest shear thinning and then an abrupt shear thickening at the critical stress. This behavior resembles in some respect that of the CPCl-NaSal solution at 40 mmol studied in Wheeler *et al.* [166]. In addition, the shear-thickening transition occurs simultaneously to the growth of a shear-induced structure, starting at the inner cylinder of the Couette (as in the shear banding transition). Because the apparent viscosity is an increasing function of $\dot{\gamma}$, the shear-induced structures in CTAB-NaSal and TTAA-NaSal solutions were described as gel phases. No connection with the phenomenology of the I-N transition was made in these systems. Other observations, such as homogeneous nucleation, fracture and re-entrant flow curves were also described for these systems [208-212].

In conclusion, although the nonlinear rheology of semidilute wormlike micelles is dominated by stress plateaus and shear banding, non standard behaviors may be found. Exotic rheologies are encountered essentially at low concentrations and for systems with binding counterions and unscreened electrostatics (see also [213-216] for further instances of non standard behaviors).

## 2.3    Theories and Interpretations

The interest of theoreticians for the rheology of wormlike micelles has come from two different sides. One group of approaches has focused on the investigation of the mechanical signature of shear banding. This includes the regimes of stress plateaus or quasi-plateaus and the slow kinetics of the stress. The mechanical theories originally establish a formal analogy between the instability shown by wormlike micelles and that observed in polymer melts in extrusion, the so-called "spurt" effect [15]. McLeish and Ball suggested in 1987 that the "spurt" instability in polymers melts would be the consequence of a multivalued non monotonic flow curve (Fig. 22) [217]. In Fig. 22, the stress increases linearly with increasing $\dot{\gamma}$, passes through a maximum (A) and then through a minimum (B) before increasing further. Two branches, one at low and one at high shear rates are separated by an mechanical instable regime (AB) characterized by a negative slope for $d\sigma/d\dot{\gamma}$. If a controlled strain experiment is performed in the instable region, the system will demix into two macroscopic bands of low ($\dot{\gamma}_1$) and high shear rate ($\dot{\gamma}_2$). Fig. 23 displays schematically the demixion between an isotropic and a nematic phases. This scheme mimics the photographs of the gap obtained by flow birefringence. In a simple shear experiment, mechanical stability requires that the shear stress is constant throughout the fluid. In the plateau region, this stress is $\sigma_P$ and so :

$$\sigma_P = \eta_1 \dot{\gamma}_1 = \eta_2 \dot{\gamma}_2 = \text{cste}$$
$$\bar{\bar{\gamma}} = (1-x)\dot{\gamma}_1 + x\dot{\gamma}_2 \tag{15}$$

where $\eta_1$ and $\eta_2$ are the viscosities in each band ($\eta_1 > \eta_2$) and x is the proportion of the solution that has been transformed. Cates and coworkers have used the reptation-reaction model developed for the linear rheology to derive a constitutive equation, and they have found a non monotonic behavior similar to that in Fig. 22 [218]. This model predicts the occurrence of a stress plateau at $\sigma_P = 2G_0/3$ and reduced rate $\dot{\gamma}\tau_R = 2.6$, values that are in qualitative agreement with experiments. The main difficulty in the early models by McLeish, Ball, Cates and coworkers on polymer melts and on semidilute micelles was to find the mechanism for the selection of the stress at which the system demixes [174,217-219]. In the unstable region, these models usually predict hysteresis and values of stresses which depend on the shear scenarios.

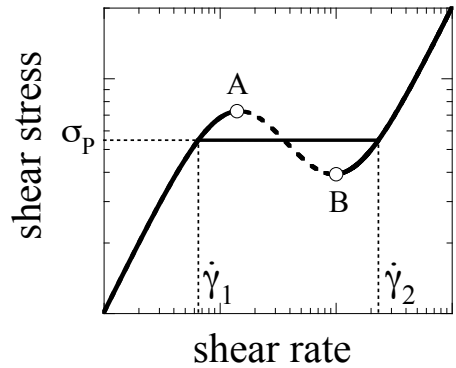

**Figure 22** : Non monotonic constitutive equation assumed for wormlike micelles. The two stable branches at low and high shear rates are separated by an unstable portion [AB]. In the plateau region, the micellar solution demixes in bands of shear rate $\dot{\gamma}_1$ and $\dot{\gamma}_2$.

The second theoretical approach is based on thermodynamical arguments. It consists in studying the non equilibrium phase behaviors of an isotropic fluid at the vicinity of the isotropic-to-nematic transition. The scheme followed by Hess, Olmsted, Dhont and coworkers started with a dispersion of rigid rods [220-225], either in the melt form or in solution using Onsager-type formalism for determining the interactions between particles [115]. For thermotropics, the temperature is the control parameter and for lyotropics it is the concentration. Imposing a shear flow to the disordered phase was shown to modify the interaction potential between rods (essentially through alignment) and triggers the transition to a nematic long range orientational state at lower temperature or concentration. The early works have predicted the flow velocity-temperature phase diagram and quantified the influence of a shear rate on the transition [223]. The thermodynamic approach underlines an important issue which does not appear in the mechanical theories, but matches more closely the data on wormlike micelles [225-227]. Above the upturn at high shear rate ($\dot{\gamma} > \dot{\gamma}_2$), the stable branch in the $\sigma(\dot{\gamma})$-curve is that of a nematic phase [117]. A major difficulty in this type of approaches concerns the validity of equilibrium thermodynamic concepts for non equilibrium or forced transitions (see [224,225,228] for discussions). There is also a conceptual difficulty in generalizing results obtained on dispersions of rods to entangled meshes of extremely long micelles. In the previous sections dedicated to experiments, we have seen the that shear banding was interpreted either in terms of a mechanical instability or in terms of a shear-induced transition, depending for instance if the solution belongs to the semidilute or the concentrated regime. It is interesting to note here that theories have faced the same duality. This duality in the interpretations of shear banding –mechanical instability *versus* shear-induced transition– has been the focus of the research on micelles during the past decade.

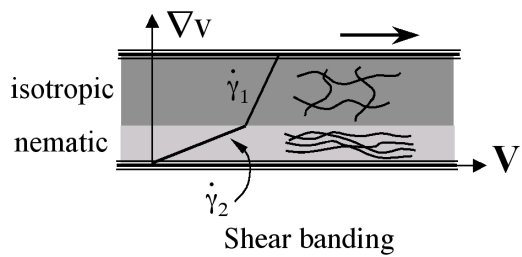

**Figure 23** : Schematic representation of the shear banding for system undergoing a shear-induced isotropic-to-nematic transition.

All theoretical approaches have tried to account for at least two experimental observations :

> *i)* the existence and robustness of the stress plateau or quasi-plateaus. As shown in many systems, there is no hysteresis in the steady state values of the stress.
> *ii)* the coexistence of macroscopic bands of different shear rates and orientations. As in an equilibrium phase transition, the proportions of the bands vary as the shear rate is changed in the plateau.

Other interesting issues related to the shear banding transition were also treated theoretically. We can cite the role of the flow geometry on the banding structure [174,175], the hypothesis of a stress-concentration coupling component [228-232] and the difference of nonlinear response between stress and strain controlled rheology [189,224,228,230-233]. A major improvement in the theoretical description was achieved by taking into account explicitly in the conservation equations an additional stress generated at the interface between bands [175,219,226,228,234-236]. This is a diffusion term which corrects the stress anomaly at the interface. For this purpose, we follow the reasoning held by Dhont in a recent paper [224]. Let us assume that, as a result of the non monotonic constitutive equation, the micellar fluid demixes into two bands when it is sheared in the unstable region. The shear rate $\dot{\gamma}(y,t)$ as well as the local viscosity $\eta(y,t)$ will then vary as function of the coordinate y along the velocity gradient direction. Even at steady state, these local quantities can also be functions of the time. A generalization of Eq. 15 at steady state (and still without the diffusion term) would give :

$$\sigma_P = \eta\left[\dot{\gamma}(y,t)\right]\dot{\gamma}(y,t) = cste \qquad (16a)$$

$$\overline{\overline{\gamma}} = \int_0^e \dot{\gamma}(y,t)dy \qquad (16b)$$

In the case of the non monotonic stress-rate relationship, the equality in Eq. 16a can not be fulfilled. We give an example of two band state in Fig. 24 and show that at the interface the stress exhibits strong spatial variations. This is due to the fact that at the interface the shear rate explores the instability region AB of the flow curve. Several authors [224,234,235] have suggested to re-write Eq. 16a as :

$$\sigma_P = \eta\left(\dot{\gamma}(y,t)\right)\dot{\gamma}(y,t) - D\left(\dot{\gamma}(y,t)\right)\frac{\partial^2\dot{\gamma}(y,t)}{\partial y^2} = cste \qquad (17)$$

A convenient choice of the diffusion coefficient $D(\dot{\gamma}(y,t))$ allows to recover the constancy of the stress throughout the gap. Once the diffusion term is introduced in the conservation equation, it is then possible to demonstrate that the stress selection rule at the plateau is robust and without hysteresis. This approach was carried out by several authors using different computing and simulations techniques. Note that it seems to be globally equivalent here to consider a non local term with a second derivative of the stress or of the shear rate [175,227,228,232]. Dhont suggests finally the shear banding transition can be assimilated to an hydrodynamic instability. The thermodynamic character of this instability results from a formal analogy between the Navier-Stokes equation and a Cahn-Hilliard equation of motion for the density in unstable systems [224,228].

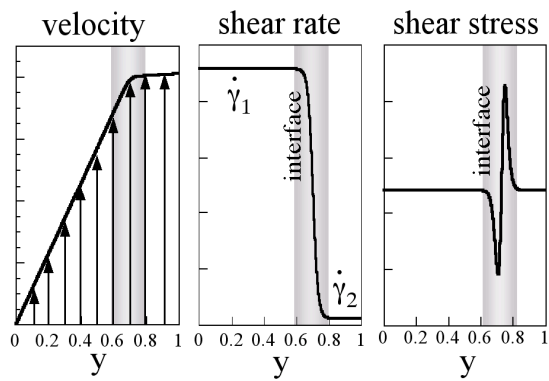

**Figure 24** : Velocity, shear rate and shear stress profiles calculated under the assumptions of shear banding and of non monotonic flow curve. In order to remove the spurious oscillations in the shear stress, a gradient term is needed (Eq. 17). This figure is taken from [224] .

## CONCLUSION

In this review, we have explored the experimental and to a lesser extent the theoretical results accumulated during the last decade on the structure and on the rheology of wormlike micelles. The systems considered in the survey are aqueous solutions of surfactants, but some results are clearly generalized to other self-assembled phases of anisotropic structures [85,86,89,90,92,93,237-240]. Our approach consisted in showing the existence of a common rheological behavior shared by most semidilute and concentrated wormlike micellar solutions. Our conclusions are summarized below.

*i)* We have shown that the local morphology of the cylindrical aggregates such as their ability to break and recombine and their flexibility play a very important role on their structure and rheology. Wormlike micelles are analogues of polymers as far as their structural properties are concerned. Their dynamical properties however are very specific and mainly dictated by the reversible breaking mechanisms. A correlation between the existence of nematic phases at high concentrations and the semiflexibility of the aggregates has been established experimentally.

*ii)* Based on a review of the surfactants and counterions that self-assemble into wormlike colloids, we have also shown that a "standard" rheological behavior can be defined. The features of this common behavior can be formulated as follows. Semidilute and concentrated solutions are viscoelastic fluids characterized by a single relaxation time. They are, to a good approximation Maxwellian fluids. In steady shear, the fluids undergo a shear banding transition associated with a plateau in the stress *versus* shear rate curves. For concentrated solutions, the new phase exhibits long range orientational order of nematic type. Because of this unique behavior, wormlike micelles are often considered as a reference system among complex fluids. Several viscoelastic systems are showing stress plateaus closely resembling those of micelles, such as dispersions of F-actin microtubules [237] or solutions of entangled DNA molecules [238]. For synthetic polymers, we can cite side-chains liquid crystals [239,240] and block copolymer microemulsions [241]. Note that except in the early work by Buxbaum *et al.* [237], explicit comparisons with surfactant micelles were made in the aforementioned works.

*iii)* There exists by now reliable models for the linear and nonlinear rheology of wormlike micelles. For the linear rheology, there is the reptation-reaction kinetics model which takes into account the two major relaxation modes shown by these colloids. As for the shear banding transition, we have followed a recent approach that identifies the transition to an hydrodynamic instability. This instability takes place because there is a non monotonic stress *versus* shear rate relationship that describes the mechanical response of the fluid (such as in Fig. 22). It is also suggested that the non monotonic flow curve originates from the existence of a nematic phase at high concentration. In this respect, the hydrodynamic instability and the shear-induced isotropic-to-nematic transition are intertwined phenomena.

*iv)* Although the rheology of entangled wormlike micelles is dominated by stress plateaus and shear banding, there exists a number of systems that do not display such behaviors. Transitions and instabilities with features different from that of shear banding are found at low concentrations and for charged systems with strongly binding counterions, such as in CPCl-NaSal and CTAB-NaSal. This raises the question of the appropriateness of the existing models for these fluids. Our point here is to outline the need for further theoretical treatments in such surfactants and other low-molecular weight systems. We think that the remarkable behaviors revealed in the nonlinear rheology (and although the linear response is Maxwellian) are indications of the existence of relaxation modes other than reptation and reversible breaking.

***Acknowledgements :***
I would like to thank all my colleagues with who I have worked during these years. It is a pleasure to acknowledge J. Appell, W. Burghardt, J.-P. Decruppe, R. Gamez-Corrales, S. Lerouge, F. Molino, P. Olmsted, G. Porte, O. Radulescu, D.C. Roux, C. Schmidt, T. Thiele and L. Walker for the fruitful and exciting discussions we had on this topic. I would like to thank also all those with who I had the opportunity to discuss about the rheology of micelles. I am personally grateful to Sandra Lerouge for having read the first draft of the manuscript and for her suggestions. The Laboratoire Léon Brillouin (Saclay, France), the Institute Laue-Langevin and the European

Synchrotron Radiation Facilities (Grenoble, France) are also acknowledged for their technical and financial supports. I have benefited from the stimulating environment provided by the Groupement de Recherche 1081 "Rhéophysique des Colloïdes et Suspensions" (CNRS funding). The research described in this review is funded in most part by the Centre National de la Recherche Scientifique in France, and in part by the European TMR-Network "Rheology of Liquid Crystals", contract number FMRX-CT96-0003 (DG 12 - ORGS).